\documentclass[useAMS,usenatbib]{mn2e}
\usepackage[english]{babel}
\usepackage{graphicx}

\title[The relationship between X-ray emission 
	  and accretion in AGNs]{Studying the relationship between X-ray emission 
	  and accretion in AGNs using the XMM-{\it Newton} Bright Serendipitous Survey}
	            
\author[R. Fanali et al.]
  {R.~Fanali,$^{1,2}$
  A.~Caccianiga,$^1$ P.~Severgnini,$^1$ R.~Della Ceca,$^1$
  E.~Marchese,$^2$ 
  \newauthor 
  F.J.~Carrera,$^3$ A.~Corral,$^4$ S.~Mateos$^3$\\
  $^1$INAF-Osservatorio Astronomico di Brera, via Brera 28, I-20121 Milan, Italy \\
  $^2$Universit\'a degli Studi di Milano Bicocca, Piazza Della Scienza 3, 20126 Milano, Italy \\
  $^3$Instituto de Fisica de Cantabria (CSIC-UC), Avenida de los Castros, 39005 Santander, Spain \\
  $^4$National Observatory of Athens (NOA), Palaia Penteli, 15236, Athens, Greece \\}

\begin{document}

\label{firstpage}

\maketitle

\begin{abstract}
{We study the link between the X-ray emission in radio-quiet AGNs and the accretion rate on the central Supermassive Black Hole using a statistically well-defined and representative sample of $71$ type $1$ AGNs extracted from the XMM-Newton Bright Serendipitous Survey. We search and quantify the statistical correlations between some fundamental parameters that characterize the X-ray emission, i.e. the X-ray spectral slope, $\Gamma$, and the X-ray ``loudness'',  and the accretion rate, both absolute ($\dot{M}$) and normalized to the Eddington luminosity (Eddington ratio, $\lambda$). We parametrize the X-ray loudness using three different quantities: the bolometric correction, K$\rm_{bol}$, the two-point spectral index $\alpha_{OX}$ and the disk/corona luminosity ratio. We find that the X-ray spectral index depends on the normalized accretion rate while the ``X-ray loudness'' depends on both the normalized and the absolute accretion rate. The dependence on the Eddington ratio,
in particular, is probably induced by the $\Gamma-\lambda$ correlation. The two proxies usually adopted in the literature to quantify the X-ray loudness of an AGN, i.e. K$\rm_{bol}$ and $\alpha_{OX}$, behave differently, with K$\rm_{bol}$ being more sensitive to the Eddington ratio and $\alpha_{OX}$ having a stronger dependence with the absolute accretion. The explanation of this result is likely related to the different sensitivity of the two parameters to the X-ray spectral index.}
\end{abstract}

\begin{keywords}
galaxies: active -  galaxies: nuclei - quasars: general - X-rays: galaxies
\end{keywords}

\maketitle

\section{Introduction}

The engine of Active Galactic Nuclei (AGNs) is powered by the accretion of matter onto the Supermassive Black Hole (SMBH), placed in the center of the host galaxy: the matter is heated ($10^5$-$10^6$ K) through viscous and magnetic processes and forms an accretion disk around the SMBH emitting in the UV-optical region. A fraction of energy is also emitted in the
X-ray band with a spectrum that can be represented, at first order, by a power-law from $0.1$ to $100$ keV at rest-frame.  In the now accepted disk-corona model (Haardt $\&$ Maraschi $1991$), the X-rays are produced in a hot (T=$10^8$-$10^9$ K) corona,
reprocessing the primary UV-optical emission of the disk via inverse-Compton mechanism. X-rays are a probe of accretion since they are produced in the very inner part of the nucleus and carry direct information on the physics very close to the SMBH. For instance, the hard X-ray spectral index ($\Gamma$) gives direct information about the energy distribution of the electrons in the corona, while the intensity of the X-ray emission with respect to the UV-optical emission quantifies the relative importance between disk and corona. This latter quantity is often parametrized with the X-ray bolometric correction K$\rm_{bol}$ (e.g. \citealt{VF2009}), defined as the ratio between bolometric luminosity and $2-10$ keV luminosity, or with the two-point-spectral index $\alpha_{OX}$ (e.g. \citealt{Vignali}), defined between 2500\AA\ and 2 keV. The different values of X-ray spectral index and of the disk/corona luminosity ratio observed from source to source are likely a consequence of fundamental differences in the physical parameters of the central engine. 

First studies, essentially based on {\itshape ROSAT} data, suggested correlations between the ``soft'' spectral index $\Gamma_{(0.5-2.4)\rm{keV}}$ and the Full Width at Half Maximum (FWHM) of H$\beta$ emission line coming from the Broad Line Region (BLR, \citealt{Wang96}, \citealt{Laor}, \citealt{SMD}, \citealt{Grupe2004}). Assuming that BLR dynamics is directly dependent on the black hole mass, this correlation was suggesting a direct link between $\Gamma_{(0.5-2.4)\rm{keV}}$ and some physical parameters like the black hole mass or accretion rate. In particular, it was suggested that the main physical driver of this correlation is the accretion rate normalized to the Eddington luminosity\footnote{The Eddington luminosity is a theoretical limit beyond which the accretion process stops for effect of radiation pressure} (Eddington ratio): sources accreting close to the Eddington limit produce the steepest values of $\Gamma_{(0.5-2.4)\rm{keV}}$ (\citealt{Wang}, \citealt{Laor}, Sulentic et al. 2000, \citealt{Grupe2004}). However, since the measured value of $\Gamma_{0.5-2.4keV}$ can be significantly contaminated by the presence of a spectral component called ``soft excess''\footnote{The ``soft excess'' is an excess of counts, at energies below 2 keV, with respect to the power-law component fitted at higher energies (typically  between 2 and 10 keV).}, it was difficult to establish on a firm ground whether it was the slope of the primary emission that correlates with the accretion rate or, instead, it was the 
intensity of the soft excess.

Using {\itshape ASCA} observations, \cite{Brandt} and \cite{Wang} have found that also the ``hard'' spectral slope ($\Gamma_{(2-10)\rm{keV}}$) has a strong dependence with the $FWHM(H\beta)$. Since the 2-10 keV energy range is not affected by the ``soft excess'', this result was considered as a compelling indication that the slope of the primary component of the X-ray emission actually correlates with $FWHM(H\beta)$. First studies made with {\itshape XMM-Newton}, {\itshape Chandra} and {\itshape Swift-XRT} have further suggested the possible presence of a second trend, i. e. an anti-correlation between $\Gamma_{(2-10)\rm{keV}}$ and the black hole mass $M \rm_{BH}$ (\citealt{Porquet}, \citealt{Piconcelli}). The availability of hard X-ray data from {\itshape XMM-Newton} and {\itshape Chandra} and of statistical relations that allow the systematic computation of $M \rm_{BH}$ on large numbers of AGNs have produced in the very recent years a big leap forward on this kind of study, extending the analysis on significantly larger samples, including up to a few hundreds of sources (\citealt{Kelly}, \citealt{GC},  \citealt{Shemmer}, \citealt{RYE}, \citealt{ZZ}, \citealt{Grupe2010}). These studies seem to confirm the presence of a correlation between the hard $\Gamma$ and the Eddington ratio (\citealt{Grupe2010}, Risaliti et al. 2009) with some exceptions (\citealt{Bianchi}). \cite{Shemmer} have also demonstrated that the observed strong anti-correlation usually observed between $\Gamma$ and $FWHM(H\beta)$ is a secondary correlation induced by the dependence between $\Gamma$ and the Eddington ratio.

Also the bolometric correction is expected to be related to the physical parameters that regulate the accretion mechanism. A possible dependence of the K$\rm_{bol}$ with the luminosity has been suggested (\citealt{Marconi}, \citealt{Hopkins}), but more recent observations seem to point out that the principal dependence is between K$\rm_{bol}$ and the Eddington ratio (\citealt{VF2007}, \citealt{Kelly}, \citealt{VF2009}, \citealt{Lusso}).
An alternative way to study the relative intensity between disk and corona is through the $\alpha\rm_{OX}$, defined as the slope between $2500\dot{A}$ and $2$ keV. Past studies generally found a strong correlation between $\alpha\rm_{OX}$ and L$\rm_{UV}$ (e.g. \citealt{Vignali}, \citealt{Marchese}) or L$\rm_{bol}$ (\citealt{Kelly}, \citealt{Shemmer}) while a dependence of $\alpha_{OX}$ with the Eddington ratio is usually weak or absent (\citealt{YER}), contrary to what has been found for K$\rm_{bol}$. This is quite surprising since K$\rm_{bol}$ and $\alpha_{OX}$ are both supposed to be proxies of the disk/corona relative intensity and therefore, they are somehow expected to behave in a similar way.

In this paper we investigate the link between X-ray properties and the accretion rate by analyzing a well defined sample of type $1$ AGNs selected from the XMM-Newton Bright Serendipitous survey (XBS). In particular, we study the spectral index $\Gamma$ estimated in the energy
range $0.5-10$ keV and $2-10$ keV and three different parameters that quantify the ``X-ray 
loudness'' i.e. the bolometric correction K$\rm_{bol}$, the $\alpha_{OX}$ and the 
disk/corona luminosity ratio (that is the ratio between the accretion disk luminosity and the $0.1-100$ keV X-ray luminosity). 
The approach followed in this study is to search for statistically significant correlations between these parameters and the value of
accretion rate, both absolute and normalized to Eddington luminosity.

The structure of the paper is the following: in Section $2$ we describe the survey, the sample selection and the parameters used for our work; in Section $3$ we describe the statistical analysis used to find the correlations between the parameters, taking into account a number of potential biases. In Section $4$ we present our results. Finally, in Section $5$ we report the summary and conclusions.

We assume here a flat $\Lambda$CDM cosmology with $H_0 = 65$ km s$^{-1}$ Mpc$^{-1}$, $\Omega_\Lambda = 0.7$ and $\Omega_M = 0.3$.

\section{XMM-Newton Bright Serendipitous survey}
   
The XMM-Newton Bright Serendipitous survey (XBS survey) is a wide-angle ($\sim$ 28 sq. deg) high Galactic latitude ($|b| > 20^\circ$) survey based on the XMM-Newton archival data. 
It is composed of two flux-limited samples: the XMM Bright Source Sample (BSS, 0.5-4.5 keV band, 389 sources) and the XMM Hard Bright Source Sample (HBSS, 4.5-7.5 keV band, 67 sources, with 56 sources in common with the BSS sample), having a flux limit of $\sim$ 7 $\times$ 10$^{-14}$ erg cm$^{-2}$ s$^{-1}$ in both energy selection bands. Selection criteria and properties of these samples are described in \cite{Della}.
The XBS is composed of sources 
that are detected serendipitously in the field-of-view of the XMM-{\it Newton} pointing, thus excluding the
targets of the observations. For this reason the XBS can be considered as representative of the X-ray
sky down to its flux limit.

To date, the spectroscopic identification level has reached 98$\%$ and 100$\%$ in the BSS and the HBSS samples, respectively. Most of the spectroscopic identifications are presented and discussed in Caccianiga et al. (2007, 2008). 

The availability of good XMM-Newton data for the sources in the XBS sample, spanning the energy range between $\sim$ 0.3 and $\sim$ 10 keV, allowed us to perform a reliable X-ray spectral analysis for almost every AGNs of the sample (\citealt{Corral}).

\subsection{The sample}

\begin{figure}
   \centering
   \includegraphics[width=6cm]{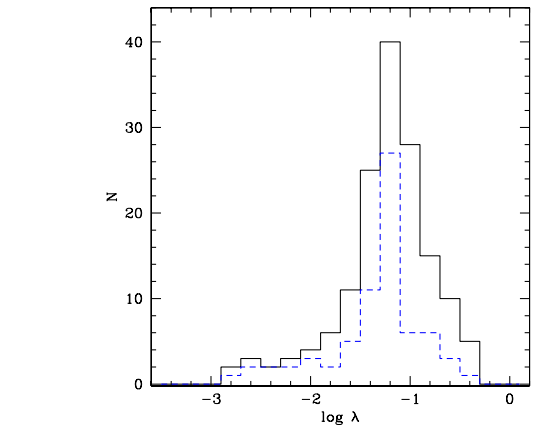}
   \caption{Eddington ratio distribution for the total sample presented in 
Marchese et al. (2012) (solid black line, $154$ AGNs) and for the 
sub-sample used here (dashed blue line, $71$ AGNs). The K-S test gives a 
probability for the null hypothesis (i.e. the two distributions are drawn 
from the same parent population) of  0.12.}
 \label{fig:histoEdd}
   \end{figure}

\begin{figure}
   \centering
   \includegraphics[width=6cm]{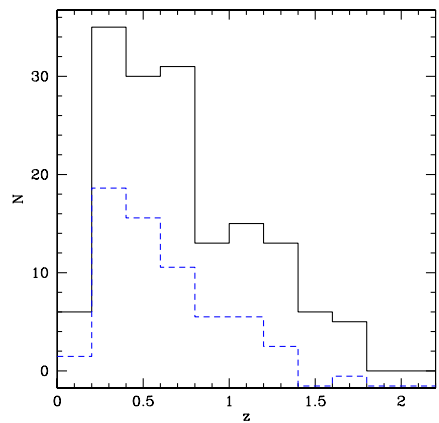}
   \caption{Redshift distribution for the total sample presented in Marchese 
et al. (2012) (solid black line, $154$ AGNs) and for the 
sub-sample used here (dashed blue line, $71$ AGNs). The K-S test 
gives a probability for the null hypothesis (i.e. the two distributions 
are drawn from the same parent population) of 0.35.}
   \label{fig:histoZ}
   \end{figure}

Since the goal of this paper is the study of the possible dependence of $\Gamma$, K$\rm_{bol}$, $\alpha_{OX}$ and the disk/corona luminosity ratio on the accretion rate, we restrict the analysis to the sub-sample of radio-quiet $154$ type $1$ AGNs for which all these parameters have been already derived by fitting the UV-optical Spectral Energy Distribution (SEDs) of the sources (\citealt{Marchese}) and by studying the X-ray and optical spectra. The radio-loud AGNs of the sample (see \citealt{Galbiati}) were not considered to avoid possible contamination from the relativistic jet to the SED. The analysis of the SEDs  was carried out on a subset of objects for which optical and UV data are available (either a detection or an upper limit) from existing catalogues (Sloan Digital Sky Survey, SDSS and Galaxy Evolution Explorer, GALEX). Since the availability of these data depends mainly on the position of the source in the sky and not on its intrinsic properties, this subset can be confidently considered as a representative sub-sample of the original one (see \citealt{Marchese}).
In addition, in order to minimize the uncertainties on the values of L$\rm_{bol}$, we have further restricted the analysis on a sub-sample of objects for which the possible effects of absorption are negligible, i. e. type $1$ AGNs with an intrinsic absorbing column density, measured from the X-ray spectra, below $5 \times 10^{20}$ cm$^{-2}$. Finally, we have excluded from the analysis the small fraction ($\sim$8\%) of ``elusive'' type $1$ AGNs, i.e. those sources whose optical spectrum is dominated by the host galaxy (see \citealt{Severgnini}, \citealt{Caccianiga2007}), due to the impossibilty of computing the BH mass through the Single Epoch spectral method (SE, e.g. see Peterson 2010 and
Marziani \& Sulentic 2012). In total, the final sample contains $71$ objects. A Kolmogorov-Smirvov test indicates that this subsample is not statistically different (at 95\% confidence level) from the original one from what concerns the Eddington ratio (Fig.~\ref{fig:histoEdd}) and the redshift (Fig.~\ref{fig:histoZ}) distributions.
 We have also evaluated the possible impact of the exclusion of ``elusive'' AGNs from the analysis (see Section $3$). The final sample used in this work consists of type $1$ AGNs with rest-frame $2-10$ keV luminosities ranging from $6 \times 10^{41}$ erg s$^{-1}$ to $9 \times 10^{46}$ erg s$^{-1}$ and redshift from $0.04$ and $2$.

\subsection{Parameters}

In this section, we describe the methods adopted to determine the parameters of interest (all reported in Table~C1).

\begin{itemize}
\item L$\rm_{bol}$ and $\dot{M}$: bolometric luminosities were obtained as the sum of the accretion disk luminosity (L$\rm_{disk}$) and the $0.1-100$ keV X-ray luminosity. L$\rm_{disk}$ was obtained by fitting the optical-UV data with a disk model (\citealt{Marchese}), while L$\rm_X$ was obtained by extrapolating the results obtained in the $2-10$ keV energy range analyzing the XMM-Newton data (\citealt{Corral}). 
As described in Marchese et al. (2012), the uncertainties on the bolometric luminosities take into 
account both the stastical errors on photometry and additional sources of
error due to the correction for the intrinsic extinction and the
long term variability (since the used photometric data are not simultaneous).

From bolometric luminosities we estimate the absolute accretion rate, defined as 
\begin{equation}
\dot{M} = \frac{L\rm_{bol}}{\eta c^2}
\end{equation}
where $\eta$ is the efficiency of the mass to energy conversion, assumed to be $0.1$. 
The uncertainties associated to the values of $\dot{M}$ in Table~C1 are those related to the bolometric
luminosity i.e. we do not assume any error on $\eta$. The uncertainty on this value is
difficult to assess. Marconi et al. (2004) estimate a range of values
for $\eta$ between 0.04 and 0.16 and, therefore, an additional uncertainity on $\dot{M}$ up 
to a factor $\sim$2 could be expected, besides that reported in Table~C1 
.
We note that, as explained above, the bolometric luminosities include the X-ray emission. Therefore, by using these bolometric luminosities to compute $\dot{M}$ we are implicitly assuming that the energy budget carried by the X-ray emission is directly related to the accretion process. 

\item $M\rm_{BH}$ and Eddington ratio $\lambda$: black hole masses of the XBS type~1 AGNs are computed in \cite{Caccianiga2013} using the Single Epoch (SE) method (Peterson 2010 and
Marziani \& Sulentic 2012). 
This method assumes that the Broad Line Region (BLR) is gravitationally influenced by the SMBH, so the virial theorem can be applied. The velocity dispersion is derived from the broad emission line widths while the BLR size is estimated from the continuum luminosity. The choice of emission lines used for $M\rm_{BH}$ estimate depends on the redshift of the source. In this sample we used H$\rm \beta$ (for $0 < z \leq 0.8$) and $MgII$ at $2798\dot{A}$ lines (for $0.8 < z \leq 2$). 
In particular, we adopted the relation discussed in  Vestergaard \& Peterson (2006) for the H$\beta$:

\begin{equation}
Log M_{BH} = 6.91 + 2 Log \frac{FWHM (H\beta)}{1000 km/s} + 0
.50 Log \frac{\lambda L_{5100\AA}}{10^{44} erg/s}
\end{equation}

and the relation presented in Shen et al. (2011) for the MgII$\lambda$2798\AA\ line:

\begin{equation}
Log M_{BH} = 6.74 + 2 Log \frac{FWHM (MgII)}{1000 km/s} + 0.62 
Log \frac{\lambda L_{3000\AA}}{10^{44} erg/s}
\end{equation}

the latter equation has been obtained by Shen et al. (2011) in such a way 
that the zero-order point (the virial factor) is the same as in the 
H$\beta$ relation presented above so that the masses are
consistently derived from these two equations (see the discussion in Shen et al. 2011). 
In both relations, the line widths refer to the broad component, and it 
is assumed that a narrow component has been subtracted during the fitting procedure and that the iron emission has been taken into account. 
All the details on how the FWHM of the emission lines have been computed are given in \cite{Caccianiga2013}. The monochromatic luminosities at 5100\AA\ ($L_{5100\AA}$) and 3000\AA\ ($L_{3000\AA}$) respectively are derived from the SED-fitting presented in \cite{Marchese}. 

The SE method is intrinsically affected by a large uncertainty, usually estimated between $0.35$ and $0.46$ dex (\citealt{Park}), essentially due to the unknown geometry of the BLR. Since the presence of large uncertainties can reduce significantly the strength of the correlations involving black hole masses (and the derived quantities) we have estimated the impact of these errors on the analysis presented here (see Section $3.2$). 

From the black hole masses we can estimate the accretion rate normalized to Eddington luminosity, defined as 
\begin{equation}
\lambda = \frac{L\rm_{bol}}{L\rm_{Edd}},
\end{equation}
where L$\rm_{Edd}$ is the Eddington luminosity:
\begin{equation}
L\rm_{Edd} = \frac{4 \pi G c M\rm_{BH} m\rm_{p}}{\sigma \rm_{e}} = 1.26 \cdot 10^{38} \left(\frac{M\rm_{BH}}{M_{\odot}}\right) \rm erg \cdot \rm s^{-1}.
\end{equation}

\item $\Gamma$, L$\rm_{(2-10)keV}$, K$\rm_{bol}$, $\alpha_{OX}$ and disk/corona luminosity ratio: the values of $\Gamma_{(0.5-10)\rm{keV}}$ and L$\rm_{(2-10)keV}$ are taken from the spectral X-ray analysis presented in \cite{Corral}. The bolometric corrections and the values of $\alpha_{OX}$ are available from \cite{Marchese}. 
In particular, the bolometric correction is defined as:

\begin{equation}
K\rm_{bol}=\frac{L\rm_{bol}}{L\rm_{(2-10)keV}}
\end{equation}

while $\alpha_{OX}$ is defined as:

\begin{equation}
\alpha_{OX} = \frac{Log(f_o/f_x)}{Log(\nu_o/\nu_x)}
\end{equation}

where $f_o$ and $f_x$ are, respectively, the rest frame monochromatic fluxes at 
$\nu_o$=1.20$\times$10$^{15}$ Hz (corresponding to $\lambda_o$=2500\AA) and
$\nu_x$=4.84$\times$10$^{17}$ Hz (corresponding to E=2 keV).

Finally, the disk/corona luminosity ratios, defined as the ratio between the accretion disk luminosity, $L_{disk}$, and the 0.1-100 keV X-ray luminosity ($L_X$), 
are computed on the basis of the luminosities presented, again, in the \cite{Marchese} work. 
\end{itemize}

%{\bfseries All the quantities used in this work are reported in Table~C1}

\section{Statistical analysis}

We perform a non parametric Spearman rank test on each correlation between X-ray properties (spectral index $\Gamma$, K$\rm_{bol}$, $\alpha_{OX}$, disk/corona luminosity ratio) and accretion rate (absolute $\dot{M}$ and normalized to Eddington luminosity, $\lambda$). When the correlation is statistically significant, we perform a fit to the data (using both the ordinary least-squares, OLS, and the bisector methods, \citealt{Isobe}) to derive the functional dependence. We define a {\itshape very significant correlation} if the probability of null hypothesis (the two quantities are not correlated) is $P \leq 0.10\%$, a {\itshape significant correlation} if $P \leq 1.00\%$ and a {\itshape marginal correlation} if $P \leq 5.00\%$. For convenience, the main correlation coefficients and probabilities computed in 
this paper are summarized in Tab.~$1$.
During the analysis, we evaluate the impact of some possible biases that we detail in the following subsections.

\subsection{Flux limited nature of the sample}

The XBS is a flux limited sample. The strong $L-z$ correlation, induced by the presence of a flux limit, may create spurious correlations or cancel real ones. This is not a problem for the correlations involving the X-ray loudness (K$\rm_{bol}$, $\alpha_{OX}$ and disk/corona luminosity ratio) since we find that these parameters are not dependent on $z$ (See Table~1). On the contrary, the values of $\Gamma$ turned out to be marginally dependent on $z$ (see Section~4.1) and, therefore, the correlations involving this quantity are potentially affected by the aforementioned problem. To exclude this possible effect, we use the partial correlation analysis (Kendall $\&$ Stuart 1979, see also Appendix B) which allows to evaluate the correlation between two parameters excluding a third variable on which both parameters depend (in this case, the redshift).
As further check of the effect of $z$ on the correlations, we analyse the correlations involving $\Gamma$ in a relatively narrow bin of z ($0 \leq z < 0.4$).

\subsection{Error impact on correlation coefficient}

As explained above, some parameters, like the black hole mass and $\lambda$ are characterized by uncertainties comparable with their variance.  This clearly reduces the strength of a correlation by decreasing the values of the correlation parameters. Under the hypothesis of independent errors, and if the average error on the quantities is known, it is possible to have an estimate of the intrinsic correlation parameter using the following relation:
\begin{equation}
r_{\rm i}= r_{\rm obs} \sqrt{\left(1 + \frac{\epsilon_x^2}{\sigma_x^2}\right)\left(1 + \frac{\epsilon_y^2}{\sigma_y^2}\right)}
\end{equation} 
where $\epsilon\rm_x$, $\epsilon\rm_y$ are the average errors on the two variables, $\sigma\rm_x^2$ and $\sigma\rm_y^2$ are the intrinsic variances on the two variables, $r_{\rm obs}$ is the observed coefficient and the term under square root is the correction factor. The intrinsic variances can be obtained from the observed variances, $\sigma\rm_{x,o}^2$ and $\sigma\rm_{y,o}^2$, by subtracting quadratically the errors, i.e.: $\sigma\rm_x^2=\sigma\rm_{x,o}^2-\epsilon\rm_x^2$ and $\sigma\rm_y^2=\sigma\rm_{y,o}^2-\epsilon\rm_y^2$.

The relation (8) can be derived from linear correlation coefficient, assuming independent errors on variables. 
Using Montecarlo simulations we have verified that it can be also applied to Spearman coefficients in case of a non-linear relation (Appendix A).

The correction presented above is particularly important for the correlations involving the Eddington ratio, since its computation is based  on the highly uncertain black-hole mass estimate. 
In this work we assume an intrinsic uncertainty on the black-hole mass of $0.40$ dex which corresponds to a correction factor for the Eddington ratio of about $\sim 1.57$.

We note that the correction discussed above can be used only to have an estimate of the intrinsic 
strength of the correlation under study. The probability associated to the correlation coefficient (to assess
the actual presence of a correlation), instead, is still the one associated to the value of $r_{\rm obs}$. 
Therefore, we will apply this correction only to the correlations that have been established to be 
statistically significant on the basis of the probabilities associated to the values of $r_{\rm obs}$.

\subsection{Induced correlations}

$\dot{M}$ and $\lambda$ are inter-related quantities since they both depend on bolometric luminosity. A possible correlation, e.g. between $\Gamma$ and $\lambda$, can create an unreal correlation between $\Gamma$ and $\dot{M}$. To verify this situation, we use partial correlation analysis which allows to calculate the correlation degree between the parameters of X-ray emission and $\lambda$, excluding the dependence on $\dot{M}$ and {\it vice-versa}. If the correlation disappears by excluding the dependence on the other variable, it is possible that the observed correlation is just induced by the other variable. Conversely, if the correlation remains, then both the observed correlations are likely to be real and not induced by the other variable.

\subsection{Elusive AGNs}
As already mentioned, we have excluded from the analysis a number of type $1$ AGNs 
whose optical spectrum is dominated by the light from the host galaxy. As discussed in 
\cite{Severgnini} and \cite{Caccianiga2007} these sources appear in the optical 
band as ``normal'' (i.e. non active galaxies) because the nuclear light is diluted by the
light coming from the host-galaxy. The spectrum shows no emission lines (the
so-called XBONG sources) or few emission lines that do not allow the clear 
recognition of the AGN and to derive the correct spectral classification. 
In \cite{Caccianiga2007} we have used the X-ray spectral analysis to assess the actual presence of the AGN
and to characterize it as ``type $2$'' (absorbed, N$_H>$4$\times$10$^{21}$ cm$^{-2}$) or 
``type $1$'' (un-absorbed, N$_H<$4$\times$10$^{21}$ cm$^{-2}$) AGN. As expected, the frequency 
of ``elusive'' AGNs is higher in type $2$ AGNs, since the absorption makes the dilution more 
effective to hide the AGN. However, also a fraction ($\sim$8\%, see Caccianiga et al 2007) 
of type $1$ AGNs 
is affected by this problem and this fraction increases rapidly when we consider type $1$ AGNs
of lower and lower X-ray luminosity, becoming very high ($>$50\%) for L$_{(2-10)keV}$ 
lower than 10$^{43}$ erg s$^{-1}$. 
In the sample considered here, i.e. the XBS type $1$ AGNs from \cite{Marchese} with low
values of N$_{H}$, there are $7$ elusive AGNs that we have excluded from the analysis. Even if
few, these objects could in principle change the results of the statistical analysis if they
are not randomly distributed. We know, for instance, that these objects typically have the
lowest values of the optical-to-X-ray flux ratio i.e. the lowest values of K$\rm_{bol}$ and
the ``flattest'' values of $\alpha_{OX}$ (all but one have $\log$ K$\rm_{bol}<$1.3 and 
$\alpha_{OX}>$-1.4). 
In order to evaluate the impact of the exclusion of these objects from the analysis, we have
derived a rough estimate of the black-hole mass using the absolute magnitude in the K-band
and adopted the relation discussed in \cite{Graham}:

\begin{equation}
\log M_{BH} = -0.37 (K+24) +8.29 
\end{equation}

where $M_{BH}$ is given in units of solar masses and K is the absolute K-band magnitude. 
We have then estimated the values of Eddington ratio and $\dot{M}$. As expected, these objects
have low accretion rates with respect to the rest of the sample ($\log\lambda<$-1.7 and 
$\log\dot{M}<$-1.3). We found that the elusive AGNs in general follow the trends observed in the total sample, so their impact on the analysis is not important. However, during the analysis presented in the following sections we will discuss, case by case, the effect of introducing the elusive AGNs on the correlation parameters. 

\section{Results}
\subsection{Spectral index $\Gamma$}
\begin{figure}
   \centering
   \includegraphics[width=6cm,angle=270]{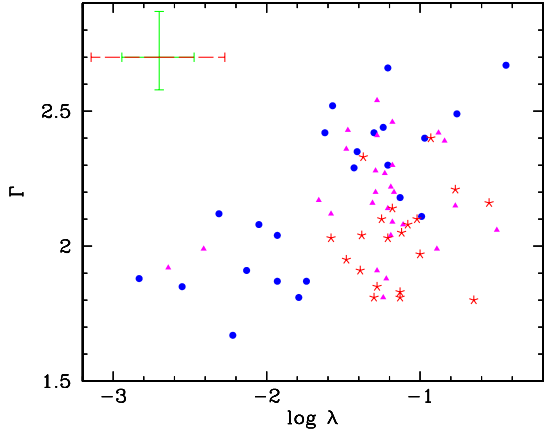}
   \caption{Plot of $\Gamma$ against $\lambda$. A typical error is shown in the upper left corner: the green solid error bar is the statistical error, the red dashed one corresponds to the total error on $\lambda$ (which includes the uncertainty related to the virial method used to estimate the black hole masses). The filled points (blue in the colour version) are sources with $0 \leq z < 0.4$, triangles (magenta in the colour version) are sources with $0.4 \leq z < 0.8$ and the stars (red in the colour version) are sources with $0.8 \leq z < 2$.}
   \label{fig:Edd}
   \end{figure}

\begin{figure}
   \centering
   \includegraphics[width=6cm,angle=-90]{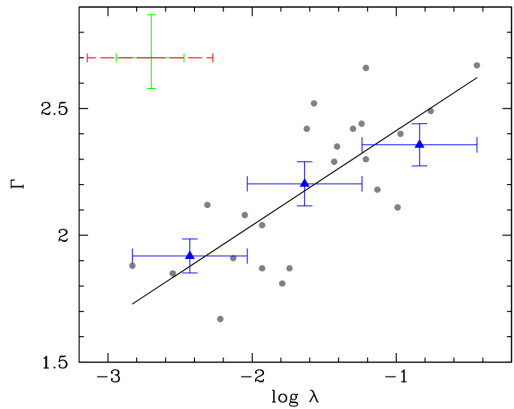}
   \caption{Plot of $\Gamma$ against $\lambda$ in the range $0 \leq z < 0.4$. A typical error is shown in the upper left corner: the green solid error bar is the statistical
   error, the red dashed one corresponds to the total error on $\lambda$ (which includes the uncertainty related to the virial method used to estimate the black hole masses). 
   The solid line represents the OLS best fit relation. Blue triangles are the binned data.}
   \label{fig:EddBIN}
   \end{figure}

The spectral index $\Gamma$ is found to marginally correlate with the Eddington ratio (r$\rm_{obs} = 0.27$, 
$P = 1.64\%$, Fig.~\ref{fig:Edd}) while the correlation between $\Gamma$ and $\dot{M}$ is not significant (r$\rm_{obs} = 0.17$, 
$P = 15.86\%$). Since $\Gamma$ marginally depends also on z (r$\rm_{obs} = -0.27$, $P = 1.64\%$) it is important
to verify whether the observed $\Gamma$-$\lambda$ correlation is in some way influenced by the luminosity-z correlation 
induced by the flux-limited nature of the sample (see discussion in Section~3.1). 
In Fig. \ref{fig:EddBIN} we present the $\Gamma-\lambda$ correlation for sources between $0 \leq z < 0.4$. This is the range that contains the greatest number of object and offers the widest coverage of $\Gamma-\lambda$ plane at the same time. The correlation in this bin of z is highly significant (r$\rm_{obs} = 0.71$, $P < 0.10\%$).

To further check this correlation, we have used the partial correlation method to exclude the dependence on z from the analysis on the total sample of $71$ AGNs. Again, we find a significant correlation with r$\rm_{obs} = 0.36$ ($P = 0.10\%$). We conclude that the $\Gamma$-$\lambda$ correlation is not induced by z. Rather, the effect of z is to weaken the correlation (see Fig.~\ref{fig:Edd}). 

It is interesting to establish the origin of the $\Gamma$-z dependence. 
The spectral index $\Gamma$ was computed using data in the range between $0.5$ and $10$ keV at rest-frame. 
In this energy range the X-ray spectrum could be contaminated by the presence of the soft-excess component.
The origin of this component is still unclear. The classical interpretation of the soft excess is represented by the high-energy tail of black body emission of the disk accretion (\citealt{CE}, \citealt{Grupe2010}). However, this interpretation was questioned when several studies showed that the observed temperature of resulting black body is remarkably constant across orders of magnitude of luminosities and BH masses (\citealt{GD}, \citealt{Crummy}). 
In the spectral analysis discussed in \cite{Corral} the soft excess component has been included in the model only if statistically required by the fit. This means that, if the number of counts is not large enough, the presence of the soft excess could be undetected and, thus, not included as additional component in the fitting procedure. In these cases the fit is expected to produce a steeper value of $\Gamma$. 
Notably, the influence of this component depends on $z$: with increasing $z$, the soft excess is confined to lower energies and it becomes negligible for $z > 1-2$ (\citealt{Mateos}, \citealt{Scott}). 
Therefore, the presence of the soft excess can produce a spurious anti-correlation between $\Gamma$ and $z$ making steeper values of $\Gamma$ at low redshifts. In order to test whether the soft excess is at the origin of the observed $\Gamma$-$z$ dependence, we
have re-computed the values of $\Gamma$ by restricting the data to energies above $2$ keV (rest-frame) in order to exclude the possible contamination due to the soft excess. 
The resulting values of $\Gamma\rm_{(2-10)keV}$ are poorly determined due to the low statistics in the hard part of the spectrum. Nevertheless, they can be used as an independent test of our conclusions. 
We find that the values of $\Gamma \rm_{(2-10)keV}$ do not depend on $z$ (r$\rm_{obs} = -0.13$, $P = 28.92\%$), 
while they depend on $\lambda$, although with a lower significance (r$\rm_{obs} = 0.24$, $P = 4.14\%$) 
when compared to $\Gamma$. In principle, given the larger errors on  $\Gamma\rm_{(2-10)keV}$ if compared to $\Gamma$, we do expect any correlation to be weaker when considering this parameter. Using equation (8) discussed in Section~3.2, it is possible to have an estimate of the impact of the larger errors on the
correlations. Since the average error on  $\Gamma \rm_{(2-10)keV}$ ($\epsilon\sim$0.20) is a factor $\sim$2.5 larger than the average error on $\Gamma$ ($\epsilon\sim$0.08) we expect 
a decrease by a factor $\sim$1.3 of the correlation coefficient just due to the increased errors. 
Thus, if $\Gamma \rm_{(2-10)keV}$ had the same dependence on z and $\lambda$ as $\Gamma$ (r$_{\rm obs}$=$-$0.27 and 
r$_{\rm obs}$=0.36, respectively) we should expect to observe correlation coefficients reduced by a factor 1.3 i.e. r$_{\rm obs}$=$-$0.21 and r$_{\rm obs}$=0.28 respectively. While the observed coefficent for the  
$\Gamma \rm_{(2-10)keV}$-$\lambda$ correlation (0.24) is quite close to the expected one (0.28), the $\Gamma \rm_{(2-10)keV}$-z correlation coefficient ($-$0.13) is nearly half than the expected one ($-$0.21).  
We consider this as an indication that the $\Gamma \rm_{(0.5-10keV)}$-$\lambda$ and $\Gamma$-$\lambda$ correlation 
has probably a similar strength while the dependence of the hard spectral index with redshift is much weaker (if any).  
These results support both the idea that the dependence between $\Gamma$ and z is (mainly) induced by the presence of the
soft excess and the idea that it is the spectral index of the primary X-ray component, and not the soft excess intensity, that correlates with the Eddington ratio. Clearly, better quality spectra, in 
particular at energies above 2 keV, are required to put these conclusions on a firmer ground.

Both $\Gamma$ and, in particular, $\lambda$ are characterized by  uncertainties that are on average large with respect to the variance of the parameters. As explained in Section $3.2$, the presence of such large errors reduces significantly the measured strength of the correlation, i.e. the value of r. In order to have a better estimate of the actual level of correlation between $\Gamma$ and $\lambda$, we have thus applied the corrections described in Section $3.2$ finding a {\itshape corrected} value of r$\rm_{i}$ of $0.6$. In case of linear correlation, the square of r$\rm_{i}$ gives an indication of how much of the observed variance on $\Gamma$ is regulated by the value of $\lambda$. We thus conclude that about $40\%$ of the variance on the spectral index is explained by $\lambda$. This is the strongest correlation found in the sample. We have evaluated the impact of the elusive AGNs (Section
$3.4$) by adding these objects to the sample. We find that their addition 
improves the $\Gamma-\lambda$ correlation while the $\Gamma-\dot{M}$ correlation 
remains not significant. We conclude that the observed $\Gamma-\lambda$ 
correlation is not due to the exclusion of the elusive AGNs.

We compute the ordinary least-squares (OLS) fit for the correlation $\Gamma-\lambda$ and we obtain
\begin{equation}
\Gamma=  0.25\log\lambda + 2.48
\end{equation}
with an error of $\pm 0.05$ on the slope, and the bisector from which
\begin{equation}
\Gamma = 0.75\log\lambda + 3.15 
\end{equation}
with an error of $\pm 0.04$ on the slope.

\subsection{Bolometric correction K$\rm_{bol}$}
\begin{figure}
   \centering
   \includegraphics[width=6cm,angle=-90]{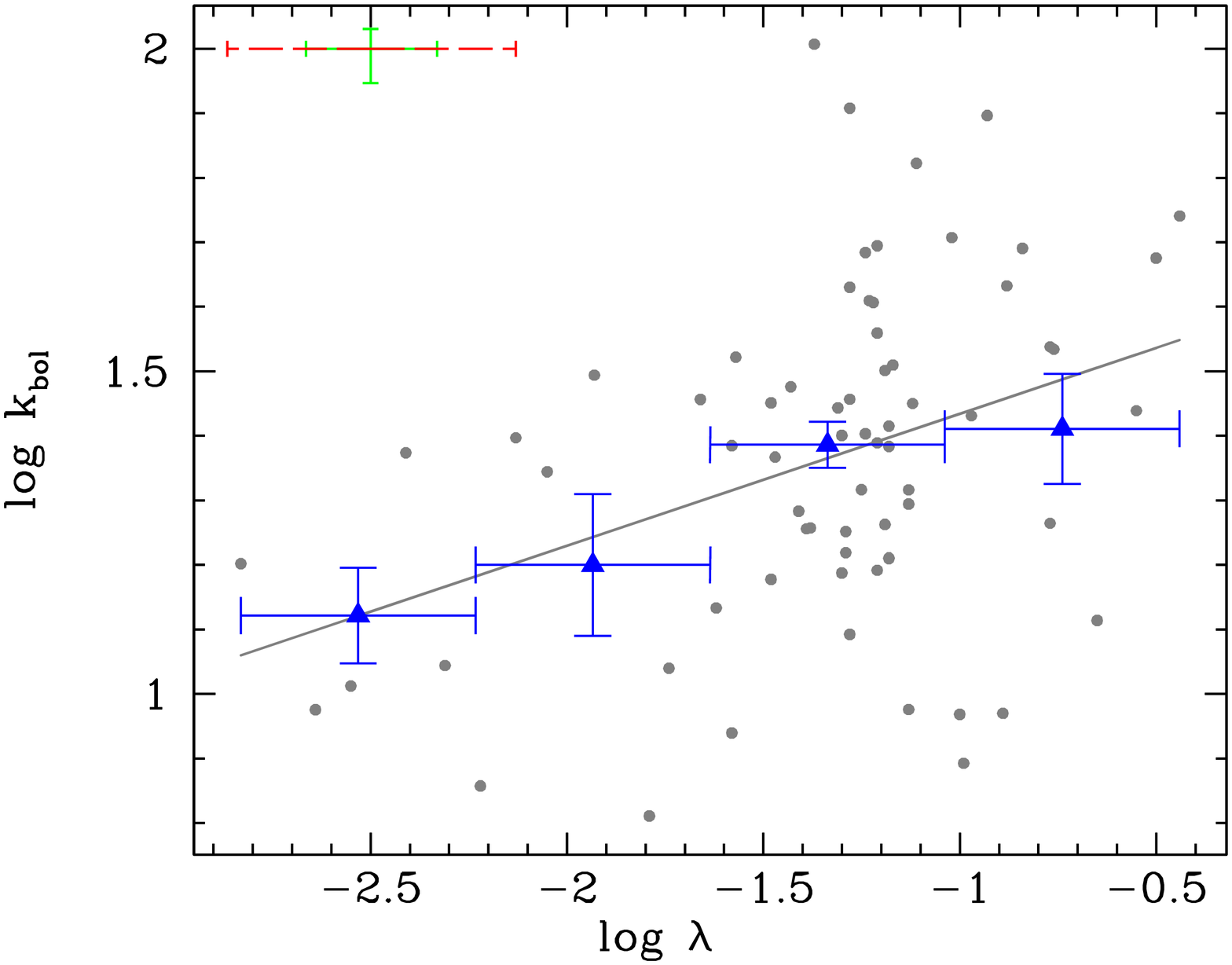}
   \caption{Plot of K$\rm_{bol}$ against $\lambda$. A typical error is shown in the upper left corner: the green solid error bar is the statistical error, the red dashed one corresponds to the total error on $\lambda$ (which includes the uncertainty related to the virial method used to estimate the black hole masses). The solid line represents the OLS best fit relation. Blue triangles are the binned data.}
   \label{fig:KbolEdd}
   \end{figure}

We find a significant correlation between K$\rm_{bol}$ and $\lambda$ (r$\rm_{\rm obs} = 0.33$, $P = 0.42\%$, Fig. \ref{fig:KbolEdd}), while the correlation between K$\rm_{bol}$ and $\dot{M}$ is only marginally significant (r$\rm_{obs} = 0.27$, $P = 2.14\%$).

By using the equation (8) to correct the correlation coefficient of K$\rm_{bol}-\lambda$ correlation, we obtain r$\rm_{i} = 0.52$ which suggests that $\sim 25\%$ of the variance on K$\rm_{bol}$ is explained by $\lambda$.
We compute the ordinary least-squares (OLS) fit for the correlation K$\rm_{bol}-\lambda$ and we obtain 
\begin{equation}
\log K\rm_{bol} = 0.18\log\lambda + 1.61
\end{equation}
with an error of $\pm 0.06$ on the slope, and the bisector from which
\begin{equation}
\log K\rm_{bol} = 0.72\log\lambda + 2.32
\end{equation}
with an error of $\pm 0.05$ on the slope.

The slope obtained using the bisector method (0.72$\pm$0.05) is in good agreement with that presented in 
\cite{Lusso} (0.75$\pm$0.04)
while the OLS slope is significantly ($\sim$2.5$\sigma$) flatter (0.18 versus 0.39). The discrepancy is
slightly reduced if we fit the data on the same range of K$\rm_{bol}$ observed in \cite{Lusso} (we 
find $0.24 \pm 0.11$). Again, we have verified that the observed correlations are not due to the exclusion of the elusive AGNs.

In conclusion, the results show that both the spectral index $\Gamma$ and the bolometric correction K$\rm_{bol}$ depend significantly on $\lambda$: steep $\Gamma$ ($\sim 2.5$) and high K$\rm_{bol}$ ($\sim 30-60$) values correspond to higher $\lambda$ ($\sim 1$), flat $\Gamma$ ($\sim 1.7$) and low K$\rm_{bol}$ values ($\sim 10$) correspond to lower $\lambda$ ($\sim 10^{-2}$). Since K$\rm_{bol}$ depends also on $\Gamma$ it is possible that the 
K$\rm_{bol}-\lambda$ correlation is induced by the (stronger) $\Gamma-\lambda$ correlation. Again, we have verified
this hypothesis using the partial correlation analysis and found that the dependence between 
K$\rm_{bol}$ and $\lambda$ can indeed be explained as induced to the $\Gamma$-$\lambda$ correlation.

\begin{table*}
\centering
\renewcommand\arraystretch{1.4} 
\begin{tabular}{|l|l|l|l|l|l|}
\hline
		 & $\Gamma$       	  & $\Gamma\rm_{(2-10)keV}$ & K$\rm_{bol}$ 	& $\alpha\rm_{OX}$ 	& $Disk/corona$  	\\
 		 & r$_{\rm obs}^{(1)}$, $P$ 	  & r$_{\rm obs}$, $P$              & r$_{\rm obs}$, $P$         	& r$_{\rm obs}$, $P$		& r$_{\rm obs}$, $P$		\\
 		 & r$_i^{(2)}$      	  & r$_i^{(2)}$             & r$_i^{(2)}$        	& r$_i^{(2)}$ 		& r$_i^{(2)}$		\\
\hline
$z$		 & $-0.27$, $1.64\%$	  & $-0.13$, $28.92\%$    & $0.03$, $80.26\%$ 	& $-0.22$, $6.29\%$    & $0.18$, $11.41\%$	\\
%                 & $-0.32$                & $-0.17$               & -                   & -                    & -                      \\
\hline
$\lambda$  	 & $0.36$, $0.10\%$   	  & $0.24$, $4.14\%$      & $0.33$, $0.42\%$	& $-0.25$, $3.32\%$	& $0.28$, $1.64\%$	\\
		 & $0.60$		  & $0.51$		  & $0.52$		& $-0.39$           	& $0.44$		\\
\hline
$\dot{M}$   	 & $0.17$, $15.86\%$  	  & 	   		  & $0.27$, $2.14\%$	& $-0.41$, $< 0.10\%$   & $0.37$, $< 0.10\%$	\\
		 & $0.19$		  & 			  & $0.24$		& $-0.41$          	& $0.37$		\\
\hline
\hline
\end{tabular}
\caption{Spearman ``rank'' correlation coefficients and probabilities for the null hypothesis for the relations discussed in the text. $^{(1)}$These values of r$_{\rm obs}$ are computed by excluding the dependence on redshift via  partial correlation. $^{(2)}$These values of r are an estimate of the ``intrinsic'' correlation coefficients computed by taking into account the role of errors (see text for details).} %$^{(3)}$This notation means that we use partial correlation to exlude the subscripted parameters.}
\end{table*}

In order to visualize these dependences we show in Fig.~$6$ two theoretical SEDs representing two extreme cases of low ($\lambda \sim 10^{-3}$, left panel) and high ($\lambda \sim 1$, right panel) accretion rate. We have built these SEDs using a Shakura-Sunyaev disk model with a maximum temperature of $3$ eV (corresponding to the average temperature of the sample sources) and a power-law in the range between $\sim 0.01$ and $100$ keV with a cut-off at $0.1$ keV. The values of the spectral index of the X-ray power-law and the relative normalizations between the disk and the X-ray component are obtained from  our $\Gamma-\lambda$ and K$\rm_{bol}-\lambda$ fits, i.e. from (10) and (12). In this way the two SEDs of Fig.~$6$ can be considered as a visual representation of the correlation analysis discussed in the previous sections. To simplify the comparison between the two SEDs, we assumed the same disk emission in both cases. It is clear from the comparison of the two SEDs that the
  variation of K$\rm_{bol}$ with $\lambda$ can be simply explained as due to a change of $\Gamma$, as suggested by the partial correlation analysis. We stress that the point where the disk emission intersects the corona emission is not fixed ``a priori'' but it comes from the values of $\Gamma$ and K$\rm_{bol}$ obtained from the fits.

\begin{figure*}
\centering
%\subfigure[]
{\includegraphics[width=7.9cm]{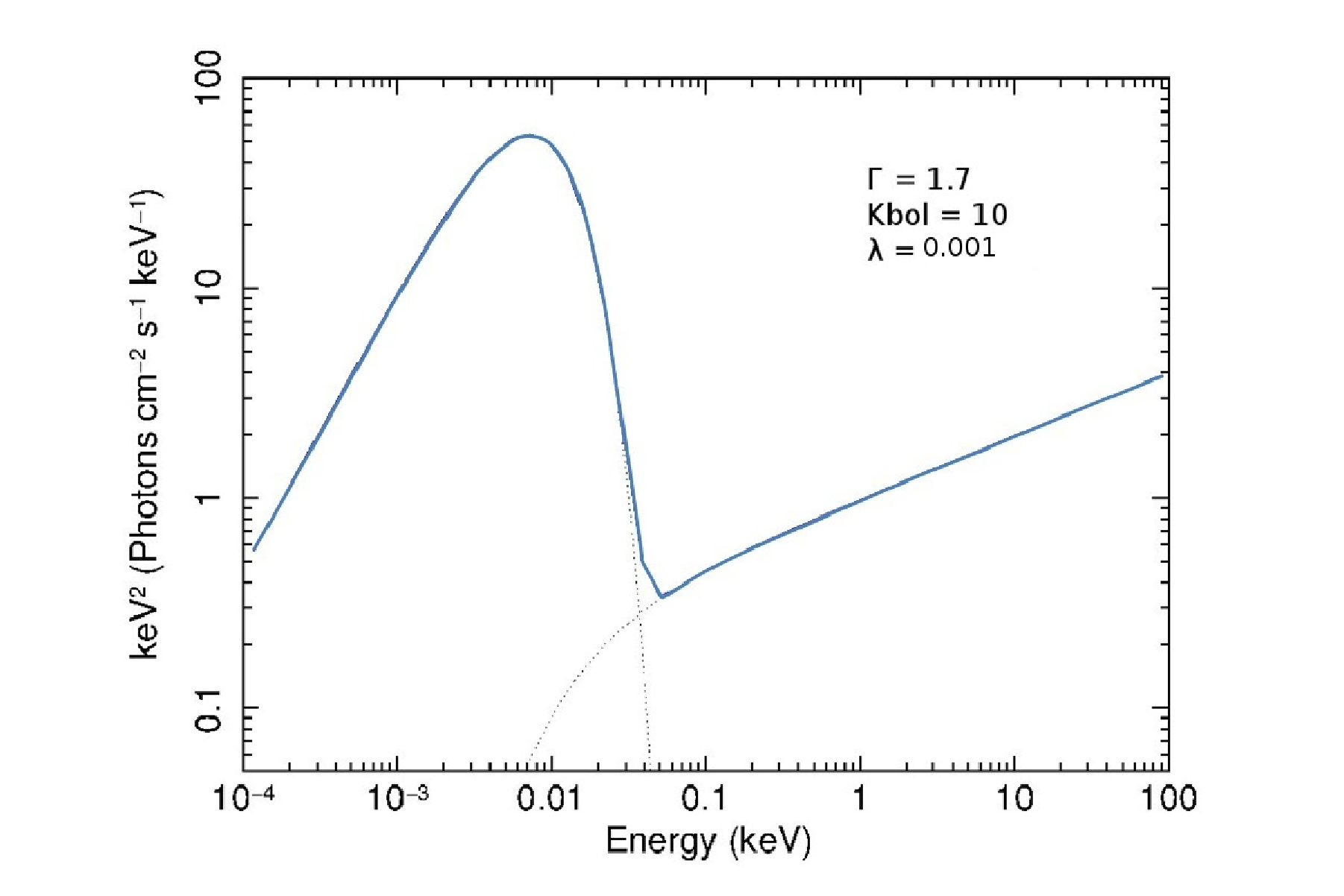}}
\hspace{5mm}
{\includegraphics[width=7.65cm]{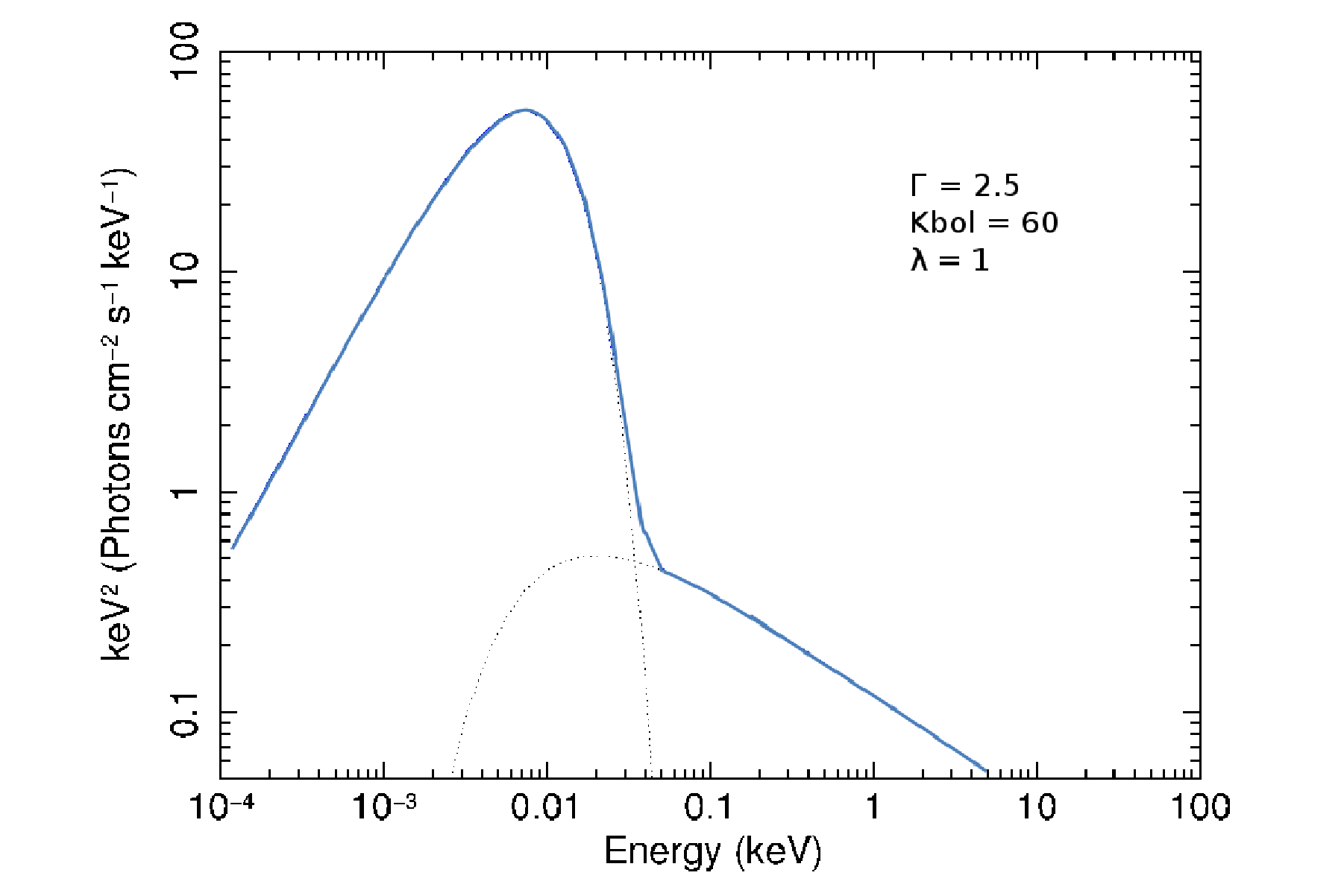}}
\caption{SEDs obtained using the results of the $\Gamma$-$\lambda$ and K$\rm_{bol}-\lambda$ best fits. The SED on the left represents the case of low accretion ($\lambda \sim 10^{-3}$): the K$\rm_{bol}$ value is low and $\Gamma$ is flat. The SED on the right represents instead the case of high accretion rate ($\lambda \sim 1$): in this case K$\rm_{bol}$ is high and $\Gamma$ is steep.}
\end{figure*}

\subsection{$\alpha\rm_{OX}$}
\begin{figure}
   \centering
   \includegraphics[width=6cm,angle=-90]{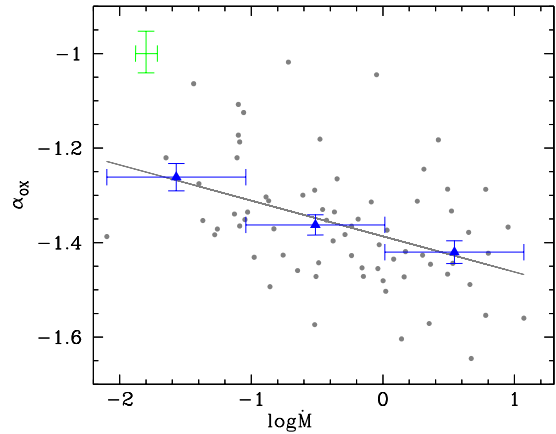}
   \caption{Plot of $\alpha \rm_{OX}$ against $\dot{M}$. A typical error is shown in the upper left corner and it is the average statistical error on $\alpha\rm_{OX}$ and $\dot{M}$. The solid line represents the OLS best fit relation. Blue triangles are the binned data.}
   \label{fig:aOXLbol}
   \end{figure}

Contrary to what is observed for the K$\rm_{bol}$, we find a marginally significant anticorrelation between $\alpha \rm _{OX}$ and $\lambda$ r$\rm_{obs}= -0.25$, $P = 3.32\%$) while we find a significant anticorrelation between $\alpha \rm_{OX}$ and $\dot{M}$ 
(r$\rm_{obs} = -0.41$, $P < 0.10\%$, Fig. \ref{fig:aOXLbol}). 
Even if we weight the correlation coefficients for the errors the dependence between $\alpha \rm_{OX} - \dot{M}$ remains the strongest one 
(r$\rm_{i} = -0.41$ versus $-0.39$).
This result confirms what is usually found in the literature i.e. that the value of $\alpha_{OX}$ 
anti-correlates with the bolometric/UV luminosity while it has weaker dependence with the Eddington ratio. The inclusion of the elusive AGNs improves the significance of both $\alpha\rm_{OX}-\lambda$ and $\alpha\rm_{OX}-\dot{M}$ correlations.

Since both K$\rm_{bol}$ and $\alpha\rm_{OX}$ are expected to be in some way proxies of the disk/corona relative intensity, the fact of finding two different dependences for these two quantities, one (K$\rm_{bol}$) on the relative accretion rate and the other ($\alpha\rm_{OX}$) on the absolute accretion, seems difficult to reconcile.
However, these two observational parameters are clearly related but not identical. The major difference is the 
fact that $\alpha\rm_{OX}$ is defined at given monochromatic frequencies while K$\rm_{bol}$ is the ratio of two
integrated quantities. For a fixed value of K$\rm_{bol}$ we can measure different values of $\alpha\rm_{OX}$ 
depending on the actual spectral shape and {\it vice-versa}. In particular, the value of $\alpha\rm_{OX}$ is 
less sensitive to the slope of the X-ray emission if compared to K$\rm_{bol}$ (r$\rm_{obs} = -0.24$, $P = 4.04\%$ for $\alpha\rm_{OX}-\Gamma$, and r$\rm_{obs} = 0.53$, $P < 0.1\%$ for K$\rm_{bol}-\Gamma$). As shown in the previous section,
the dependence of K$\rm_{bol}$ to the Eddington ratio is probably induced by a change of $\Gamma$ so it is 
probable that the weaker dependence of $\alpha\rm_{OX}$ on $\lambda$ is a consequence of the weaker dependence
of $\alpha\rm_{OX}$ on $\Gamma$.

On the other hand, the significant dependence of $\alpha_{OX}$ with $\dot{M}$ suggests that the disk/corona
relative intensity depends also on the absolute accretion rate. We test this hypothesis in the next section by studying directly the disk/corona luminosity ratio.

\subsection{Disk-corona luminosity ratio}
\begin{figure}
   \centering
   \includegraphics[width=6cm,angle=-90]{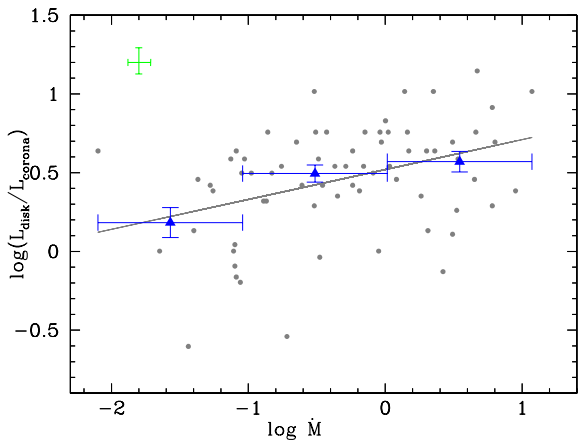}
   \caption{Plot of disk-corona luminosity ratio against $\dot{M}$. A typical error is shown in the upper left corner and it is the average statistical error on disk-corona luminosity ratio and $\dot{M}$. The solid line represents the OLS best fit relation. Blue triangles are the binned data.}
   \label{fig:DCMpunto}
   \end{figure}

\begin{figure}
   \centering
   \includegraphics[width=6cm,angle=-90]{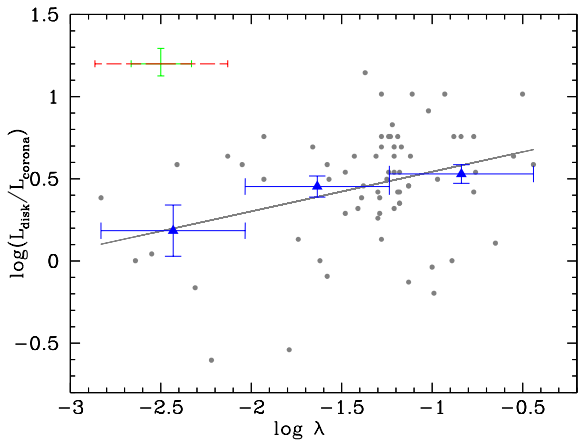}
   \caption{Plot of disk-corona ratio against $\lambda$. A typical error is shown in the upper left corner: the green solid error bar is the statistical error, the red dashed one corresponds to the total error on $\lambda$ (which includes the uncertainty related to the virial method used to estimate the black hole masses). The solid line represents the OLS best fit relation. Blue triangles are the binned data.}
   \label{fig:DCEdd}
   \end{figure}

The dependences of K$\rm_{bol}$ and $\alpha\rm_{OX}$ discussed in the previous sections seem to suggest a complex
relationship between the disk/corona luminosity ratio and the accretion. From the one hand, there is 
a significant dependence on the Eddington ratio, probably related to a change of X-ray slope with $\lambda$. 
On the other hand, there could be also a dependence of the disk/corona luminosity ratio on the absolute level of 
accretion rate. 
We now want to study directly the dependence of the disk/corona luminosity ratio with accretion. 
As expected, the situation in this case is more complex than the K$\rm_{bol}$ and $\alpha\rm_{OX}$ case. We find
significant correlation with $\dot{M}$ (r$\rm_{obs} = 0.37$, $P < 0.10\%$, Fig. \ref{fig:DCMpunto}) and a marginally significant correlation
with $\lambda$ (r$\rm_{obs} = 0.28$, $P = 1.64\%$, Fig. \ref{fig:DCEdd}). We find a similar result if we add the elusive AGNs into the analysis. The strength of the two correlations, once corrected for the errors,
is quite similar (r$\rm_{i} \sim 0.4$) so it is difficult to establish if there is a dominant correlation that explains
also the other one. It is thus possible that both correlations are in fact present i.e. that the disk/corona
relative intensity depends both on $\lambda$ and $\dot{M}$, as expected from the combination of the results obtained for K$\rm_{bol}$ and $\alpha\rm_{OX}$.

\section{Discussion and conclusions}

In this paper we studied the link between X-ray emission and accretion rate in a statistically well-defined and complete sample of $71$ type $1$ AGNs extracted from the XMM-Newton Bright Serendipitous survey (XBS). The X-ray properties analyzed here are the spectral index $\Gamma$ in the range $0.5-10$ keV and $2-10$ keV band and the X-ray ``loudness'' parametrized with both the bolometric correction K$\rm_{bol}$ (defined as the ratio between bolometric luminosity and $2-10$ keV luminosity) and the two-points spectral index $\alpha_{OX}$. We have also directly analysed the disk/corona luminosity ratio. The spectral index gives direct information about the energy distribution of the electrons in the corona, while the other 3 parameters quantify, in different ways, the relative importance between disk and corona.

We have considered different possible biases which can influence final results, such as:
\begin{itemize}
\item soft excess contamination;
\item redshift induced correlations (important in flux limited samples);
\item impact of errors on correlation coefficients (especially on $M\rm_{BH}$ estimate);
\item interconnected dependences due to the fact that the parameters considered in the analysis are not all independent;
\item the impact  of the exclusion of ``elusive'' AGNs from the analysis on the final results.
\end{itemize}

The results can be summarized as follows:
\begin{itemize}

\item the spectral index $\Gamma$ depends significantly on accretion rate normalized to Eddington luminosity; in particular, $\sim 40\%$ of $\Gamma$ variance could be explained by $\lambda$. This correlation is not due to the soft excess contamination,  but it probably reflects a true dependence of the slope of the primary X-emission with $\lambda$. 
The $\Gamma$-$\lambda$ dependence can be speculatively attributed to the effect of cooling of the electrons in the corona:  for high values of $\lambda$, a large number of photons comes from the accretion disk and cools corona electrons rapidly, thus producing steep X-ray spectra while for low values of $\lambda$, less photons are available and this makes electron cooling inefficient, thus producing flat X-ray spectra (see for instance \citealt{Cao}). 

\item the ``X-ray loudness'' depends both on $\lambda$ and $\dot{M}$ but the dependence with $\lambda$ is probably just the consequence of the (stronger) $\Gamma-\lambda$ dependence;

\item The strenght of the dependence between the ``X-ray loudness'' and $\lambda$ or $\dot{M}$ is different depending 
on whether we parametrize the X-ray loudness using the K$\rm_{bol}$
or the $\alpha\rm_{OX}$:  while K$\rm_{bol}$ seems to depend mainly on $\lambda$, the values of $\alpha\rm_{OX}$ show a stronger dependence with $\dot{M}$. The explanation is likely connected to the different sensitivity of these two parameters to the X-ray spectral index.

\end{itemize}

\section*{Acknowledgments}

We thank the referee for useful comments that improved the paper. We acknowledge Massimo Dotti, Francesco Haardt, Monica Colpi and Valentina Braito for useful discussions and  Laura Maraschi and Tommaso Maccacaro for the precious comments.
The authors acknowledge financial support from ASI (grant n. I/088/06/0), from the Italian Ministry of Education, Universities and Research (PRIN2010-2011, grant n. 2010NHBSBE) and from the Spanish Ministry of Economy and Competitiveness through grant AYA2012-31447.

\bibliographystyle{mn2e}

\newpage
\begin{appendix}
\section{Error impact on correlation coefficient}
\begin{figure}
\centering
{\includegraphics[width=7cm]{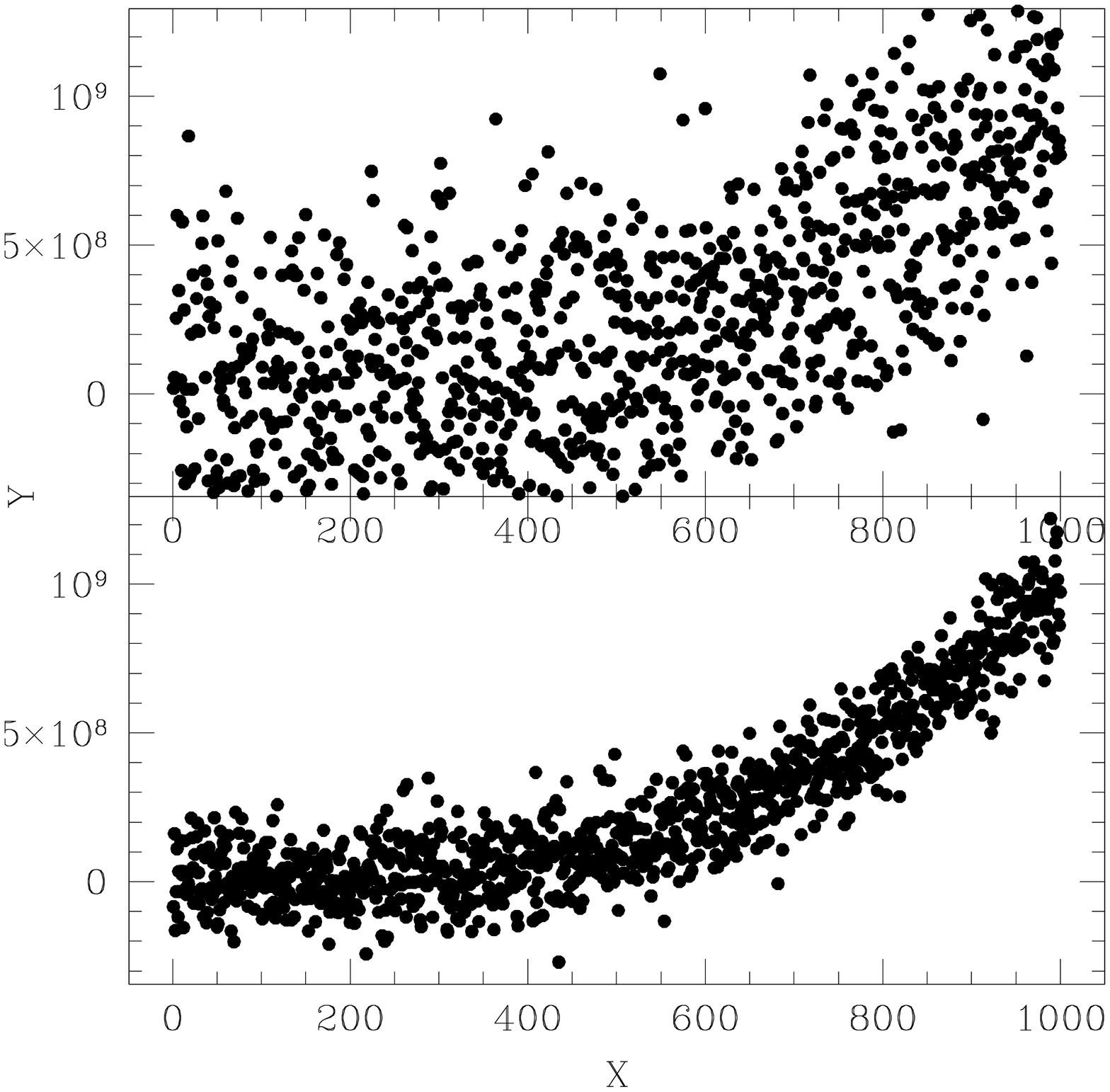}}
\caption[]{\footnotesize{Numerical simulation ($\sim$1000 points) that shows the impact of a big error (comparable with the variance of the variable, in this example Y) on the Y-X correlation. In this example we assume a correlation coefficient $r \sim 0.87$ (lower panel) and we add an error on Y comparable with the intrinsic variance on Y. The resulting correlation (upper panel) is significantly reduced ($r \sim 0.62$).}
\label{fig:3.4}}
\end{figure}

Some parameters used in this analysis (like the black hole mass and the Eddington ratio) are characterized by very large errors, principally related to the method adopted to estimate the black hole masses. If the error is comparable to the variance of a variable, this can reduce the strength of a correlation by decreasing the values of the correlation coefficients. 
We estimate the intrinsic correlation parameter by using the relation:
\begin{equation}
r_{\rm i} = r_{\rm obs} \sqrt{\left(1 + \frac{\epsilon_x^2}{\sigma_x^2}\right)\left(1 + \frac{\epsilon_y^2}{\sigma_y^2}\right)}.
\end{equation} 
where $\epsilon\rm_x$, $\epsilon\rm_y$ are the average errors on the two variables, $\sigma\rm_x^2$ and $\sigma\rm_x^2$ are the intrinsic (i.e. not folded with the errors) variances on the two variables, $r_{\rm obs}$ is the observed coefficient and the term under square root of this variable is the correction factor. This relation can be derived from linear correlation coefficient, assuming independent errors on variables. 
Using Montecarlo simulations we have verified that it can be also applied to Spearman coefficients in case of a non-linear relation.
Fig. \ref{fig:3.4} represents a Montecarlo simulation where we show the case of a cubic correlation between two variables, X and Y, with an intrinsic correlation coefficient $r \sim 0.87$ (lower panel in Fig. \ref{fig:3.4}). If we add an error on Y comparable to the variance on this variable, the coefficient correlation  is reduced to $r \sim 0.62$ (upper panel in Fig. \ref{fig:3.4}). 

\begin{figure}
\centering
\includegraphics[width=6cm]{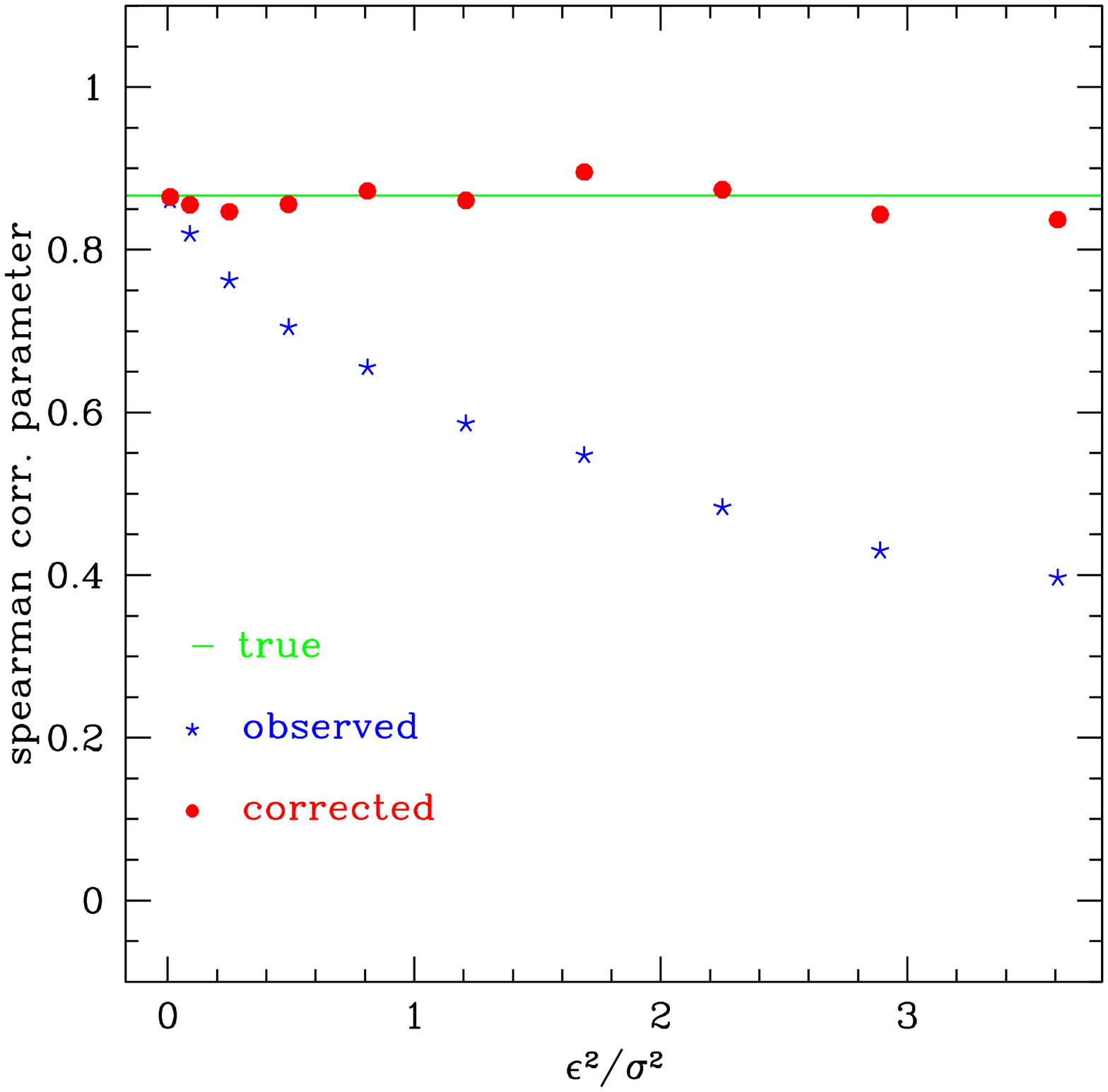}
\caption[]{Results of numerical simulations that show the variation of the observed correlation coefficient (blue stars) with respect to the error$^2$/variance ratio, assuming a starting value of $r \sim 0.87 $. The red points represent corrected r values.}
\label{fig:3.5}
\end{figure}

We repeated these simulations for different values of errors and the trend of the observed r$_{obs}$ is shown in Fig. \ref{fig:3.5} (blue stars). In Fig. \ref{fig:3.5} we also report the values of r$_i$ estimated according to equation (A1) (red points). The starting value of $r_i \sim 0.9$ is reasonably recovered.

\section{Partial correlations}
As explained in Section $3.1$, in a flux-limited sample like the XBS the luminosity is strongly correlated with redshift. This relation could give rise to spurious correlations. A way of dealing with the problem is to examine the correlations between luminosities excluding the dependence on redshift via partial correlation analysis.
If $r_{12}$ is the correlation coefficient between $x_1$ and $x_2$ and $r_{13}$ and $r_{23}$ are the correlation coefficients of the two variables with $z$, the correlation coefficient between $x_1$ and $x_2$, excluding the effect of $z$ is:
\begin{equation}
r\rm_{12,3} = \frac{r\rm_{12}-r\rm_{13}r\rm_{23}}{\sqrt{(1-r^2\rm_{13})(1-r^2\rm_{23})}}.
\end{equation}
This equation can be generalized to more than three variables. For example, in the case of four variables it becomes:
\begin{equation}
r\rm_{12,34} = \frac{r\rm_{12,4}-r\rm_{13,4}r\rm_{23,4}}{\sqrt{(1-r^2\rm_{13,4})(1-r^2\rm_{23,4})}}.
\end{equation}

\section{The sample}
In this section we present the table including all the quantities used in the analysis discussed in the text.

\begin{table*}
\label{table}
\caption{Main properties of the sample of $70$ type $1$ AGNs analysed in this work.}
\begin{tabular}{c c c c c c r c c r}
\hline\hline
name & z & $\Gamma$ & $\Gamma_{2-10}$ & LogK$_{bol}$ & LogM$_{BH}$ & Log$\dot{M}$ & Log$\lambda$ & $\alpha_{OX}$ & Log(L$_{disk}$/L$_{corona})$\\
\hline
XBSJ000027.7--250442 & 0.336 & 1.87$^{+0.06}_{-0.05}$ & 1.57$^{+0.27}_{-0.25}$ & 1.32$^{+0.09}_{-0.11}$ &
8.63$^{+0.10}_{-0.12}$ & -0.94$^{+0.09}_{-0.12}$ & -1.93$^{+0.13}_{-0.17}$ & -1.430 &  0.497  \\
XBSJ000031.7--245502 & 0.284 & 2.29$^{+0.08}_{-0.08}$ & 1.86$^{+0.42}_{-0.52}$ & 1.48$^{+0.10}_{-0.34}$ &
8.02$^{+1.32}_{-0.25}$ & -1.05$^{+0.11}_{-0.33}$ & -1.43$^{+1.32}_{-0.41}$ & -1.362 &  0.638  \\
XBSJ000102.4--245850 & 0.433 & 2.12$^{+0.08}_{-0.07}$ & 1.89$^{+0.28}_{-0.34}$ & 0.94$^{+0.07}_{-0.06}$ &
8.16$^{+0.15}_{-0.14}$ & -1.06$^{+0.07}_{-0.06}$ & -1.58$^{+0.17}_{-0.15}$ & -1.106 & -0.093  \\
XBSJ001831.6+162925 & 0.553 & 2.39$^{+0.04}_{-0.04}$ & 2.11$^{+0.14}_{-0.17}$ & 1.69$^{+0.10}_{-0.08}$ &
8.54$^{+0.06}_{-0.05}$ & 0.06$^{+0.10}_{-0.09}$ & -0.84$^{+0.12}_{-0.10}$ & -1.501 &  0.757  \\
XBSJ002618.5+105019 & 0.473 & 2.04$^{+0.04}_{-0.04}$ & 1.95$^{+0.16}_{-0.15}$ & 1.50$^{+0.10}_{-0.08}$ &
9.03$^{+0.10}_{-0.14}$ & 0.20$^{+0.10}_{-0.08}$ & -1.19$^{+0.14}_{-0.16}$ & -1.469 &  0.757  \\
XBSJ002637.4+165953 & 0.554 & 2.15$^{+0.04}_{-0.03}$ & 2.07$^{+0.13}_{-0.13}$ & 1.26$^{+0.09}_{-0.11}$ &
8.21$^{+0.11}_{-0.41}$ & -0.20$^{+0.08}_{-0.11}$ & -0.77$^{+0.14}_{-0.42}$ & -1.363 &  0.420  \\
XBSJ003418.9--115940 & 0.850 & 2.10$^{+0.27}_{-0.16}$ & 2.03$^{+0.43}_{-0.51}$ & 1.32$^{+0.14}_{-0.16}$ &
8.84$^{+0.11}_{-0.13}$ & -0.05$^{+0.14}_{-0.16}$ & -1.25$^{+0.18}_{-0.21}$ & -1.310 &  0.497  \\
XBSJ005009.9--515934 & 0.610 & 2.28$^{+0.09}_{-0.08}$ & 2.11$^{+0.44}_{-0.42}$ & 1.22$^{+0.08}_{-0.06}$ &
8.45$^{+0.35}_{-0.58}$ & -0.48$^{+0.08}_{-0.06}$ & -1.29$^{+0.36}_{-0.58}$ & -1.287 &  0.289  \\
XBSJ010432.8--583712 & 1.640 & 1.95$^{+0.05}_{-0.04}$ & 1.76$^{n.d.}_{n.d.}$ & 1.18$^{+0.10}_{-0.10}$ &
9.94$^{+0.08}_{-0.09}$ & 0.82$^{+0.10}_{-0.09}$ & -1.48$^{+0.13}_{-0.13}$ & -1.285 &  0.289  \\
XBSJ012025.2--105441 & 1.338 & 2.40$^{+0.21}_{-0.18}$ & 2.32$^{+0.36}_{-0.31}$ & 1.90$^{+0.14}_{-0.14}$ &
9.68$^{+0.08}_{-0.08}$ & 1.11$^{+0.14}_{-0.14}$ & -0.93$^{+0.16}_{-0.16}$ & -1.558 &  1.016  \\
XBSJ012119.9--110418 & 0.204 & 2.66$^{+0.23}_{-0.14}$ & 3.56$^{+1.54}_{-1.16}$ & 1.69$^{+0.12}_{-0.12}$ &
8.13$^{+0.08}_{-0.09}$ & -0.72$^{+0.12}_{-0.12}$ & -1.21$^{+0.14}_{-0.15}$ & -1.424 &  0.540  \\
XBSJ013204.9--400050 & 0.445 & 2.42$^{+0.17}_{-0.14}$ & 2.48$^{+0.52}_{-0.43}$ & 1.63$^{+0.13}_{-0.13}$ &
8.05$^{+0.13}_{-0.12}$ & -0.47$^{+0.13}_{-0.13}$ & -0.88$^{+0.18}_{-0.18}$ & -1.470 &  0.757  \\
XBSJ020029.0+002846 & 0.174 & 2.42$^{+0.10}_{-0.10}$ & 2.22$^{+0.66}_{-0.80}$ & 1.13$^{+0.06}_{-0.05}$ &
7.65$^{+0.17}_{-0.20}$ & -1.61$^{+0.06}_{-0.05}$ & -1.62$^{+0.18}_{-0.21}$ & -1.218 &  0.002  \\
XBSJ021808.3--045845 & 0.712 & 1.91$^{+0.04}_{-0.03}$ & n.d.$^{ }_{ }$ & 1.46$^{+0.10}_{-0.08}$ &
9.45$^{+0.06}_{-0.05}$ & 0.53$^{+0.09}_{-0.08}$ & -1.28$^{+0.11}_{-0.09}$ & -1.465 &  0.694  \\
XBSJ021817.4--045113 & 1.080 & 1.83$^{+0.04}_{-0.03}$ & 1.78$^{+0.08}_{-0.07}$ & 0.98$^{+0.06}_{-0.07}$ &
9.23$^{+0.07}_{-0.09}$ & 0.46$^{+0.05}_{-0.07}$ & -1.13$^{+0.09}_{-0.11}$ & -1.181 & -0.128  \\
XBSJ021820.6--050427 & 0.646 & 1.81$^{+0.04}_{-0.04}$ & 1.70$^{+0.14}_{-0.13}$ & 1.40$^{+0.06}_{-0.12}$ &
8.76$^{+0.06}_{-0.10}$ & -0.12$^{+0.06}_{-0.12}$ & -1.24$^{+0.08}_{-0.16}$ & -1.451 &  0.540  \\
XBSJ021923.2--045148 & 0.632 & 2.41$^{+0.07}_{-0.04}$ & 2.20$^{+0.23}_{-0.22}$ & 1.63$^{+0.10}_{-0.08}$ &
8.81$^{+0.07}_{-0.05}$ & -0.11$^{+0.10}_{-0.08}$ & -1.28$^{+0.12}_{-0.09}$ & -1.470 &  0.757  \\
XBSJ024200.9+000020 & 1.112 & 2.03$^{+0.05}_{-0.04}$ & 1.91$^{+0.13}_{-0.17}$ & 1.38$^{+0.07}_{-0.04}$ &
9.79$^{+0.06}_{-0.04}$ & 0.57$^{+0.07}_{-0.04}$ & -1.58$^{+0.09}_{-0.06}$ & -1.439 &  0.587  \\
XBSJ024207.3+000037 & 0.385 & 2.52$^{+0.12}_{-0.08}$ & 1.93$^{+0.31}_{-0.27}$ & 1.52$^{+0.06}_{-0.07}$ &
8.42$^{+0.10}_{-0.10}$ & -0.79$^{+0.06}_{-0.07}$ & -1.57$^{+0.12}_{-0.12}$ & -1.368 &  0.497  \\
XBSJ031015.5--765131 & 1.187 & 1.91$^{+0.02}_{-0.02}$ & 1.84$^{+0.06}_{-0.06}$ & 1.26$^{+0.09}_{-0.12}$ &
10.02$^{+0.08}_{-0.10}$ & 0.99$^{+0.09}_{-0.12}$ & -1.39$^{+0.12}_{-0.16}$ & -1.364 &  0.385  \\
XBSJ033208.7--274735 & 0.544 & 1.99$^{+0.09}_{-0.07}$ & 1.92$^{+0.19}_{-0.24}$ & 1.37$^{+0.07}_{-0.13}$ &
9.60$^{+0.07}_{-0.11}$ & -0.45$^{+0.07}_{-0.13}$ & -2.41$^{+0.10}_{-0.17}$ & -1.441 &  0.587  \\
XBSJ050446.3--283821 & 0.840 & 1.97$^{+0.11}_{-0.08}$ & 1.87$^{+0.46}_{-0.38}$ & 0.97$^{+0.08}_{-0.07}$ &
8.20$^{+0.35}_{-0.36}$ & -0.44$^{+0.08}_{-0.06}$ & -1.00$^{+0.36}_{-0.36}$ & -1.178 & -0.037  \\
XBSJ050501.8--284149 & 0.257 & 2.18$^{+0.05}_{-0.05}$ & 2.15$^{+0.39}_{-0.35}$ & 1.29$^{+0.14}_{-0.11}$ &
7.44$^{+0.11}_{-0.09}$ & -1.33$^{+0.14}_{-0.11}$ & -1.13$^{+0.18}_{-0.14}$ & -1.350 &  0.457  \\
XBSJ051955.5--455727 & 0.562 & 2.09$^{+0.04}_{-0.04}$ & 2.00$^{+0.38}_{-0.33}$ & 1.21$^{+0.08}_{-0.10}$ &
8.51$^{+0.07}_{-0.08}$ & -0.31$^{+0.08}_{-0.11}$ & -1.18$^{+0.11}_{-0.14}$ & -1.262 &  0.351  \\
XBSJ065400.0+742045 & 0.362 & 2.30$^{+0.19}_{-0.12}$ & 2.37$^{+0.60}_{-0.49}$ & 1.56$^{+0.13}_{-0.13}$ &
8.24$^{+0.10}_{-0.10}$ & -0.61$^{+0.12}_{-0.13}$ & -1.21$^{+0.16}_{-0.16}$ & -1.456 &  0.694  \\
XBSJ074352.0+744258 & 0.800 & 2.03$^{+0.07}_{-0.06}$ & 1.92$^{+0.20}_{-0.25}$ & 1.39$^{+0.09}_{-0.12}$ &
9.06$^{+0.08}_{-0.09}$ & 0.21$^{+0.10}_{-0.12}$ & -1.21$^{+0.13}_{-0.15}$ & -1.418 &  0.638  \\
XBSJ080504.6+245156 & 0.980 & 2.08$^{+0.10}_{-0.10}$ & 1.77$^{+0.32}_{-0.28}$ & 0.96$^{+0.04}_{-0.04}$ &
8.39$^{+0.14}_{-0.17}$ & -0.33$^{+0.03}_{-0.05}$ & -1.08$^{+0.14}_{-0.18}$ & -1.155 & -0.075  \\
XBSJ080608.1+244420 & 0.357 & 2.49$^{+0.04}_{-0.03}$ & 2.21$^{+0.18}_{-0.23}$ & 1.53$^{+0.06}_{-0.07}$ &
8.15$^{+0.07}_{-0.07}$ & -0.25$^{+0.06}_{-0.07}$ & -0.76$^{+0.09}_{-0.10}$ & -1.380 &  0.540  \\
XBSJ100100.0+252103 & 0.794 & 2.20$^{+0.07}_{-0.04}$ & 2.12$^{+0.17}_{-0.16}$ & 1.25$^{+0.08}_{-0.07}$ &
8.78$^{+0.06}_{-0.05}$ & -0.15$^{+0.08}_{-0.07}$ & -1.29$^{+0.10}_{-0.09}$ & -1.346 &  0.385  \\
XBSJ100309.4+554135 & 0.673 & 2.27$^{+0.07}_{-0.06}$ & 1.86$^{+0.35}_{-0.42}$ & 1.61$^{+0.07}_{-0.08}$ &
8.87$^{+0.05}_{-0.05}$ & -0.01$^{+0.08}_{-0.08}$ & -1.23$^{+0.09}_{-0.09}$ & -1.454 &  0.757  \\
XBSJ100828.8+535408 & 0.384 & 2.04$^{+0.12}_{-0.09}$ & 1.29$^{+0.64}_{-0.54}$ & 1.49$^{+0.07}_{-0.08}$ &
8.75$^{+0.30}_{-0.24}$ & -0.82$^{+0.07}_{-0.08}$ & -1.93$^{+0.31}_{-0.25}$ & -1.491 &  0.757  \\
XBSJ100921.7+534926 & 0.387 & 2.35$^{+0.08}_{-0.05}$ & 1.94$^{+0.35}_{-0.34}$ & 1.28$^{+0.08}_{-0.10}$ &
8.22$^{+0.12}_{-0.12}$ & -0.83$^{+0.08}_{-0.10}$ & -1.41$^{+0.14}_{-0.16}$ & -1.309 &  0.320  \\
XBSJ101838.0+411635 & 0.577 & 2.36$^{+0.07}_{-0.06}$ & 2.09$^{+0.30}_{-0.26}$ & 1.45$^{+0.06}_{-0.07}$ &
8.79$^{+0.05}_{-0.06}$ & -0.33$^{+0.07}_{-0.07}$ & -1.48$^{+0.09}_{-0.09}$ & -1.332 &  0.540  \\
XBSJ101850.5+411506 & 0.577 & 2.30$^{+0.05}_{-0.03}$ & 2.17$^{+0.15}_{-0.20}$ & 1.38$^{+0.06}_{-0.07}$ &
8.89$^{+0.05}_{-0.04}$ & 0.07$^{+0.07}_{-0.08}$ & -1.18$^{+0.09}_{-0.08}$ & -1.372 &  0.540  \\
XBSJ101922.6+412049 & 0.239 & 2.12$^{+0.16}_{-0.05}$ & n.d.$^{ }_{ }$ & 1.04$^{+0.05}_{-0.04}$ &
8.90$^{+0.08}_{-0.75}$ & -1.05$^{+0.05}_{-0.04}$ & -2.31$^{+0.09}_{-0.75}$ & -1.186 & -0.163  \\
XBSJ103120.0+311404 & 1.190 & 1.85$^{+0.12}_{-0.08}$ & 1.76$^{+0.20}_{-0.18}$ & 1.09$^{+0.09}_{-0.05}$ &
9.27$^{+0.09}_{-0.06}$ & 0.35$^{+0.09}_{-0.05}$ & -1.28$^{+0.13}_{-0.08}$ & -1.240 &  0.132  \\
XBSJ103154.1+310732 & 0.299 & 1.88$^{+0.13}_{-0.12}$ & 1.42$^{+0.84}_{-0.76}$ & 1.20$^{+0.06}_{-0.07}$ &
9.25$^{+0.26}_{-0.19}$ & -1.22$^{+0.06}_{-0.06}$ & -2.83$^{+0.27}_{-0.20}$ & -1.369 &  0.385  \\
XBSJ103932.7+205426 & 0.237 & 1.87$^{+0.11}_{-0.09}$ & 1.87$^{+0.63}_{-0.54}$ & 1.04$^{+0.07}_{-0.05}$ &
8.02$^{+0.17}_{-0.13}$ & -1.36$^{+0.07}_{-0.05}$ & -1.74$^{+0.18}_{-0.14}$ & -1.273 &  0.132  \\
XBSJ103935.8+533036 & 0.229 & 2.08$^{+0.15}_{-0.10}$ & 2.22$^{+0.56}_{-0.43}$ & 1.34$^{+0.09}_{-0.12}$ &
8.70$^{+0.07}_{-0.09}$ & -0.99$^{+0.09}_{-0.12}$ & -2.05$^{+0.11}_{-0.15}$ & -1.333 &  0.587  \\
XBSJ104026.9+204542 & 0.465 & 1.99$^{+0.03}_{-0.03}$ & 1.88$^{+0.13}_{-0.13}$ & 0.97$^{+0.04}_{-0.05}$ &
8.52$^{+0.05}_{-0.08}$ & -0.01$^{+0.04}_{-0.04}$ & -0.89$^{+0.06}_{-0.09}$ & -1.043 &  0.002  \\
XBSJ104509.3--012442 & 0.472 & 2.14$^{+0.11}_{-0.06}$ & 2.13$^{+0.29}_{-0.31}$ & 1.19$^{+0.06}_{-0.06}$ &
8.00$^{+0.06}_{-0.05}$ & -0.85$^{+0.05}_{-0.06}$ & -1.21$^{+0.08}_{-0.08}$ & -1.301 &  0.320  \\
XBSJ104912.8+330459 & 0.226 & 1.67$^{+0.12}_{-0.09}$ & 1.91$^{+0.46}_{-0.39}$ & 0.86$^{+0.03}_{-0.03}$ &
8.46$^{+0.21}_{-0.18}$ & -1.40$^{+0.02}_{-0.03}$ & -2.22$^{+0.21}_{-0.18}$ & -1.060 & -0.603  \\
XBSJ105014.9+331013 & 1.012 & 2.33$^{+0.37}_{-0.20}$ & 2.45$^{+0.95}_{-0.69}$ & 2.01$^{+0.10}_{-0.13}$ &
9.72$^{+0.13}_{-0.09}$ & 0.71$^{+0.10}_{-0.13}$ & -1.37$^{+0.16}_{-0.16}$ & -1.643 &  1.146  \\
XBSJ105239.7+572431 & 1.113 & 2.10$^{+0.02}_{-0.02}$ & 2.04$^{+0.12}_{-0.16}$ & 1.71$^{+0.07}_{-0.09}$ &
9.48$^{+0.05}_{-0.06}$ & 0.82$^{+0.07}_{-0.09}$ & -1.02$^{+0.09}_{-0.11}$ & -1.550 &  0.914  \\
XBSJ105316.9+573551 & 1.204 & 1.80$^{+0.02}_{-0.02}$ & 1.97$^{+0.14}_{-0.18}$ & 1.11$^{+0.05}_{-0.05}$ &
8.82$^{+0.12}_{-0.14}$ & 0.53$^{+0.05}_{-0.05}$ & -0.65$^{+0.13}_{-0.15}$ & -1.285 &  0.109  \\
XBSJ105624.2--033522 & 0.635 & 2.16$^{+0.09}_{-0.06}$ & 2.20$^{+0.26}_{-0.23}$ & 1.44$^{+0.07}_{-0.08}$ &
8.75$^{+0.05}_{-0.05}$ & -0.20$^{+0.07}_{-0.08}$ & -1.31$^{+0.09}_{-0.09}$ & -1.425 &  0.638  \\
XBSJ112022.3+125252 & 0.406 & 2.22$^{+0.09}_{-0.08}$ & 1.75$^{+0.38}_{-0.50}$ & 1.26$^{+0.06}_{-0.07}$ &
8.26$^{+0.06}_{-0.06}$ & -0.57$^{+0.06}_{-0.06}$ & -1.19$^{+0.08}_{-0.08}$ & -1.295 &  0.420  \\
XBSJ120359.1+443715 & 0.641 & 2.43$^{+0.12}_{-0.12}$ & 2.57$^{+0.40}_{-0.34}$ & 1.37$^{+0.11}_{-0.10}$ &
8.77$^{+0.06}_{-0.06}$ & -0.34$^{+0.11}_{-0.10}$ & -1.47$^{+0.13}_{-0.12}$ & -1.396 &  1.600  \\
XBSJ123116.5+641115 & 0.454 & 1.92$^{+0.05}_{-0.05}$ & 1.91$^{+0.25}_{-0.22}$ & 0.98$^{+0.04}_{-0.04}$ &
9.21$^{+0.18}_{-0.13}$ & -1.07$^{+0.05}_{-0.04}$ & -2.64$^{+0.19}_{-0.14}$ & -1.217 &  0.002  \\
XBSJ123759.6+621102 & 0.910 & 2.05$^{+0.04}_{-0.04}$ & 1.89$^{+0.12}_{-0.15}$ & 1.45$^{+0.07}_{-0.08}$ &
9.16$^{+0.05}_{-0.05}$ & 0.40$^{+0.06}_{-0.08}$ & -1.12$^{+0.08}_{-0.09}$ & -1.443 &  0.638  \\
\hline
\end{tabular}
\end{table*}
\newpage
\addtocounter{table}{-1}
\begin{table*}
\caption{continua}
\begin{tabular}{c c c c c c r c c r}
\hline\hline
name & z & $\Gamma$ & $\Gamma_{2-10}$ & LogK$_{bol}$ & LogM$_{BH}$ & Log$\dot{M}$ & Log$\lambda$ & $\alpha_{OX}$ & Log(L$_{disk}$/L$_{corona}$)\\
\hline
XBSJ123800.9+621338 & 0.440 & 2.54$^{+0.04}_{-0.05}$ & 2.01$^{+0.26}_{-0.33}$ & 1.91$^{+0.07}_{-0.09}$ &
8.44$^{+0.09}_{-0.10}$ & -0.48$^{+0.07}_{-0.08}$ & -1.28$^{+0.11}_{-0.13}$ & -1.571 &  1.016  \\
XBSJ124214.1--112512 & 0.820 & 1.81$^{+0.05}_{-0.05}$ & 1.60$^{+0.16}_{-0.15}$ & 1.32$^{+0.10}_{-0.08}$ &
8.89$^{+0.07}_{-0.06}$ & 0.12$^{+0.09}_{-0.08}$ & -1.13$^{+0.11}_{-0.10}$ & -1.431 &  0.457  \\
XBSJ124607.6+022153 & 0.491 & 2.46$^{+0.12}_{-0.08}$ & 1.81$^{+0.57}_{-0.48}$ & 1.42$^{+0.06}_{-0.07}$ &
8.40$^{+0.10}_{-0.10}$ & -0.42$^{+0.06}_{-0.07}$ & -1.18$^{+0.12}_{-0.12}$ & -1.326 &  0.420  \\
XBSJ124641.8+022412 & 0.934 & 2.21$^{+0.07}_{-0.05}$ & 2.00$^{+0.19}_{-0.23}$ & 1.54$^{+0.04}_{-0.08}$ &
9.11$^{+0.02}_{-0.06}$ & 0.70$^{+0.03}_{-0.08}$ & -0.77$^{+0.04}_{-0.10}$ & -1.485 &  0.757  \\
XBSJ124949.4--060722 & 1.053 & 2.16$^{+0.07}_{-0.06}$ & 1.70$^{+0.31}_{-0.28}$ & 1.44$^{+0.07}_{-0.08}$ &
8.53$^{+0.05}_{-0.06}$ & 0.34$^{+0.06}_{-0.08}$ & -0.55$^{+0.08}_{-0.10}$ & -1.422 &  0.638  \\
XBSJ132101.6+340656 & 0.335 & 2.44$^{+0.04}_{-0.04}$ & 2.18$^{+0.18}_{-0.20}$ & 1.68$^{+0.07}_{-0.08}$ &
8.49$^{+0.07}_{-0.08}$ & -0.39$^{+0.06}_{-0.09}$ & -1.24$^{+0.09}_{-0.12}$ & -1.351 &  0.757  \\
XBSJ133807.5+242411 & 0.631 & 2.08$^{+0.10}_{-0.08}$ & 1.84$^{+0.32}_{-0.35}$ & 1.82$^{+0.07}_{-0.09}$ &
8.93$^{+0.04}_{-0.06}$ & 0.18$^{+0.07}_{-0.09}$ & -1.11$^{+0.08}_{-0.11}$ & -1.601 &  1.016  \\
XBSJ134749.9+582111 & 0.646 & 2.20$^{+0.02}_{-0.02}$ & 1.93$^{+0.06}_{-0.06}$ & 1.51$^{+0.07}_{-0.08}$ &
9.65$^{+0.07}_{-0.07}$ & 0.84$^{+0.06}_{-0.08}$ & -1.17$^{+0.09}_{-0.11}$ & -1.419 &  0.694  \\
XBSJ140102.0--111224$^1$ & 0.037 & 1.91$^{+0.02}_{-0.02}$ & 1.74$^{+0.12}_{-0.12}$ & 1.40$^{+0.19}_{-0.35}$ &
7.71$^{+0.96}_{-0.82}$ & -2.06$^{+0.07}_{-0.09}$ & -2.13$^{+0.96}_{-0.82}$ & -1.382 &  0.638  \\
XBSJ141531.5+113156 & 0.257 & 1.85$^{+0.02}_{-0.04}$ & n.d.$^{ }_{ }$ & 1.01$^{+0.04}_{-0.05}$ &
9.13$^{+0.17}_{-0.15}$ & -1.06$^{+0.05}_{-0.05}$ & -2.55$^{+0.18}_{-0.16}$ & -1.174 &  0.043  \\
XBSJ144937.5+090826 & 1.260 & 1.81$^{+0.07}_{-0.04}$ & 1.80$^{+0.11}_{-0.10}$ & 1.19$^{+0.08}_{-0.06}$ &
9.50$^{+0.07}_{-0.06}$ & 0.56$^{+0.08}_{-0.06}$ & -1.30$^{+0.11}_{-0.08}$ & -1.332 &  0.261  \\
XBSJ160706.6+075709 & 0.233 & 2.42$^{+0.09}_{-0.08}$ & 2.02$^{+0.62}_{-0.55}$ & 1.40$^{+0.06}_{-0.07}$ &
7.70$^{+0.10}_{-0.11}$ & -1.24$^{+0.06}_{-0.07}$ & -1.30$^{+0.12}_{-0.13}$ & -1.382 &  0.420  \\
XBSJ160731.5+081202 & 0.226 & 2.67$^{+0.22}_{-0.13}$ & 2.32$^{+0.72}_{-0.87}$ & 1.74$^{+0.09}_{-0.08}$ &
6.99$^{+0.09}_{-0.11}$ & -1.09$^{+0.09}_{-0.08}$ & -0.44$^{+0.13}_{-0.14}$ & -1.335 &  0.587  \\
XBSJ165406.6+142123 & 0.641 & 1.88$^{+0.12}_{-0.08}$ & 1.93$^{+0.39}_{-0.34}$ & 1.61$^{+0.13}_{-0.13}$ &
8.90$^{+0.09}_{-0.10}$ & 0.04$^{+0.13}_{-0.13}$ & -1.22$^{+0.16}_{-0.16}$ & -1.478 &  0.829  \\
XBSJ165425.3+142159 & 0.178 & 2.11$^{+0.04}_{-0.02}$ & 1.97$^{+0.13}_{-0.13}$ & 0.89$^{+0.05}_{-0.04}$ &
7.61$^{+0.26}_{-0.36}$ & -1.02$^{+0.04}_{-0.04}$ & -0.99$^{+0.26}_{-0.36}$ & -1.124 & -0.196  \\
XBSJ165448.5+141311 & 0.320 & 1.81$^{+0.07}_{-0.04}$ & 1.78$^{+0.20}_{-0.27}$ & 0.81$^{+0.02}_{-0.02}$ &
8.75$^{+0.05}_{-0.06}$ & -0.68$^{+0.02}_{-0.02}$ & -1.79$^{+0.05}_{-0.06}$ & -1.016 & -0.540  \\
XBSJ205635.7--044717 & 0.217 & 2.40$^{+0.10}_{-0.08}$ & 1.83$^{+0.52}_{-0.73}$ & 1.43$^{+0.11}_{-0.11}$ &
7.60$^{+0.10}_{-0.09}$ & -1.01$^{+0.11}_{-0.11}$ & -0.97$^{+0.15}_{-0.14}$ & -1.347 &  0.497  \\
XBSJ213002.3--153414 & 0.562 & 2.06$^{+0.13}_{-0.12}$ & 2.31$^{+0.33}_{-0.30}$ & 1.68$^{+0.13}_{-0.14}$ &
8.53$^{+0.08}_{-0.07}$ & 0.39$^{+0.14}_{-0.13}$ & -0.50$^{+0.16}_{-0.15}$ & -1.567 &  1.016  \\
XBSJ214041.4--234720 & 0.490 & 2.17$^{+0.05}_{-0.05}$ & 1.91$^{+0.19}_{-0.24}$ & 1.46$^{+0.10}_{-0.08}$ &
9.31$^{+0.06}_{-0.06}$ & 0.01$^{+0.10}_{-0.08}$ & -1.66$^{+0.12}_{-0.10}$ & -1.400 &  0.694  \\
XBSJ225050.2--642900 & 1.251 & 2.04$^{+0.04}_{-0.04}$ & 1.93$^{+0.12}_{-0.12}$ & 1.26$^{+0.11}_{-0.11}$ &
9.71$^{+0.11}_{-0.08}$ & 0.69$^{+0.11}_{-0.10}$ & -1.38$^{+0.16}_{-0.13}$ & -1.374 &  0.457  \\
XBSJ231342.5--423210 & 0.973 & 2.14$^{+0.08}_{-0.04}$ & 2.00$^{+0.16}_{-0.15}$ & 1.21$^{+0.08}_{-0.06}$ &
9.12$^{+0.11}_{-0.11}$ & 0.30$^{+0.08}_{-0.06}$ & -1.18$^{+0.14}_{-0.13}$ & -1.309 &  0.351  \\
\hline
\end{tabular}
\addtocounter{table}{-1}
\caption{Column $1$: source name; Column $2$: redshift; Column $3$: X-ray spectral index between $0.5$ and $10$ keV; Column $4$: X-ray spectral index between $2$ and $10$ keV; Column $5$: Logarithm of the bolometric correction; Column $6$: Logarithm of the black hole mass in units of solar masses; Columns $7$: Logarithm of the absolute accretion rate in units of solar masses per year; Column $8$: Logarithm of Eddington ratio; Column $9$: two-point spectral index; Column $10$: Logarithm of the disk/corona luminosity ratio. All errors are at 68\% confidence level (please note that in 
Corral et al. 2011 the reported errors on $\Gamma$ are at 90\% confidence level). 
$^1$The X-ray luminosity of XBSJ140102.0--111224 reported here is different from the value that appears in Corral et al. (2011) because of a typo discovered in that paper. Therefore, also the derived
quantities, like K$_{bol}$, $\alpha_{OX}$ are different from what reported in Marchese et al. (2012).}
\end{table*}

\end{appendix}

\end{document}